\documentclass[aps,prd,superscriptaddress,showpacs,tighten,nofootinbib]{revtex4-1}
\usepackage{epsfig}
\usepackage{amssymb}
\usepackage{rotate}
\usepackage{float}
\usepackage{graphicx}
\usepackage{subfig}
\newcommand{\ba}{\begin{array}}
\newcommand{\ea}{\end{array}}
\newcommand{\bd}{\begin{displaymath}}
\newcommand{\ed}{\end{displaymath}}
\newcommand{\be}{\begin{equation}}
\newcommand{\ee}{\end{equation}}
\newcommand{\bea}{\begin{eqnarray}}
\newcommand{\eea}{\end{eqnarray}}

\def\a{\alpha}
\def\b{\beta}
\def\g{\gamma}

\def\e{\epsilon}

\def\m{\mu}
\def\n{\nu}

\def\n{\nu}

\def\th13 {\theta_{13}}

\usepackage{graphicx}

\catcode`\@=11
\def\lsim{\mathrel{\mathpalette\@versim<}}
\def\gsim{\mathrel{\mathpalette\@versim>}}
\def\@versim#1#2{\vcenter{\offinterlineskip
\ialign{$\m@th#1\hfil##\hfil$\crcr#2\crcr\sim\crcr } }}
\catcode`\@=12

\catcode`@=12
\begin{document}
\title{Non-standard interactions and bimagic baseline for neutrino oscillations}

\author{Rathin Adhikari}
\email{rathin@ctp-jamia.res.in}
\affiliation{Centre for Theoretical Physics, Jamia Millia Islamia (Central University),  Jamia Nagar, New Delhi-110025, INDIA}

\author{Arnab Dasgupta}
\email{arnab@ctp-jamia.res.in}
\affiliation{Centre for Theoretical Physics, Jamia Millia Islamia (Central University),  Jamia Nagar, New Delhi-110025, INDIA}

\author{Zini Rahman}
\email{zini@ctp-jamia.res.in}
\affiliation{Centre for Theoretical Physics, Jamia Millia Islamia (Central University),  Jamia Nagar, New Delhi-110025, INDIA}

\begin{abstract}
For standard interactions of neutrinos with matter bimagic baseline of length about 2540 Km is known to be suitable for getting good discovery limits of neutrino mass hierarchy, $\sin^2 \theta_{13}$ and $CP$ violation
in the $\nu_e \rightarrow \nu_{\mu}$ oscillation channel. We discuss how even in presence of non-standard interactions (NSIs) of neutrinos with matter this baseline is found to be suitable for getting these discovery limits. This is because even in presence of NSIs one could get the $\nu_e \rightarrow \nu_\mu $ oscillation probability to be almost independent of $CP$ violating phase $\delta$ and $\theta_{13}$ for one hierarchy and highly dependent on these two for the other hierarchy over certain parts of neutrino energy range. For another certain part of the  energy range the reverse of this happens with respect to the hierarchies. We present the discovery limits of NSIs also in the same neutrino energy range. However, as with the increase
of neutrino energy the NSI effect in the above oscillation probability gets relatively more pronounced in comparison to
the vacuum oscillation parameters, so we consider higher neutrino energy range also for getting better discovery limits of NSIs. 
Analysis presented here for 2540 Km could also be implemented for longer bimagic baseline $> 6000$ Km.
\end{abstract}

\maketitle
\section{Introduction}
The present experiments on neutrino oscillations confirms that there is mixing between different flavours of neutrinos ($\nu_e$, $\nu_{\mu}$, $\nu_{\tau}$). The probability of neutrino oscillations depends on various parameters of the neutrino mixing matrix-the PMNS matrix \cite{pmns}. The  current experiments tells us about two of the angles $\theta_{23}$ and $\theta_{12}$ \cite{pdg} with some accuracy  but for $\theta_{13}$ only the upper bound is given \cite{pdg} and the $CP$ violating phase $\delta$ is totally unknown. Although the mass squared difference of the different neutrinos ($\Delta m_{ij}^2=m_i^2-m_j^2$) are known to us but the sign of $\Delta m_{31}^2$ (which is related to mass hierarchy) is still unknown. Due to the correlations among these unknowns there are ambiguities \cite{ki} in analysing neutrino oscillation datas. To reduce these ambiguities one may consider neutrino oscillation experiments in long baseline \cite{base} - particulaly in magic baseline \cite{magic1}. The magic baseline  satisfies certain condition on its length from the detector and is found to be about 7500 Km where the perturbative expression of probability $P({\nu_e \rightarrow \nu_\mu})$  becomes independent of $\delta$ upto order $\alpha^2$ (where $\alpha=\Delta m^2_{21}/{\Delta m^2_{31}}$). Although this could result in finding out the other unknown oscillation parameters conveniently but for measurement of $\delta$ this baseline is not suitable. To circumvent this problem,   conditions on neutrino energy has been considered \cite{magenergy1,dighe} in ${\nu_e \rightarrow \nu_\mu}$ channel for which also the perturbative expression of probability becomes independent of $\delta$ but only on a part of the
neutrino energy spectrum. But the other part of the spectrum will be sensitive to $CP$ violating phase $\delta $.  As pointed out in \cite{dighe}, one may consider satisfying two different energy conditions simultaneously for two 
hierarchies in the same baseline  which results in fixing the length of bimagic baseline to about 2540 Km.  Unlike magic 
baseline, here the baseline is shorter so the neutrino flux for such baseline is reduced by lesser amount at the detector.  Also in this oscillation channel  ${\nu_e \rightarrow \nu_\mu}$ which has been considered in this work, the detection of muon is easier in comparison to some other channels where the detection  of electron is required. 

We have studied the effect of non-standard interactions (NSIs) of neutrinos with matter in bimagic baseline. At first there is
discussion on how to obtain the perturbative expression of the probability of oscillation  upto order $\a^2$ in ${\nu_e \rightarrow \nu_\mu}$ channel  in presence of NSIs. The NSIs  present in the $\nu_e \rightarrow \nu_\mu $ oscillation probability are  $\e_{ee} , \e_{e\mu}$ and  $\e_{e\tau}$ among which $\e_{ee} , \e_{e\mu}$ are $\lsim \a$ but $\e_{ee}$ has no such constraints 
in considering perturbation.  
In our numerical analysis we have considered the experimentally allowed range which covered the perturbative regime and also has gone beyond that.  We  have also presented the $\delta$ and $\theta_{13}$ independent perturbative expression of the oscillation probability in presence of the NSIs under two different magic energy conditions corresponding to two different hierarchies.  
In presence of NSIs different discovery limits for hierarchy of neutrino masses, for $\sin^2\theta_{13}$ and also for 
$CP$ violation have been shown in figures.  Discovery limits of NSIs particularly $\e_{ee}$, $\e_{e\mu}$ and $\e_{e\tau}$ for specific values of  $\theta_{13}$ and  $\delta$ in their allowed range have also been presented.  One may note that to satisfy the bimagic conditions one requires lower neutrino energy within 5 GeV. However, the perturbative expression of oscillation probability shows that the NSI effect
will relatively increase in comparison to other neutrino oscillation parameters in vacuum if the neutrino energy is higher. For this reason we have considered higher neutrino energy of 50 GeV also
to study the discovery limits of NSIs in the same baseline of 2540 Km for which better limits are obtained.

\section{$\nu_e \rightarrow \nu_\mu$ ocillation probability in presence of NSI}
The fermion-neutrino interaction in matter is defined by the Lagrangian:
\bea
\label{eq:Lang}
\mathcal{L}_{NSI}^{M}=\frac{G_F}{\sqrt 2}\e_{\a \b}^{f P}[\bar{\n}_{\b}\g^{\m}L\n_{\a}][\bar{f}\g_{\m}Pf]  
 + \frac{G_F}{\sqrt 2}\left(\e_{\a \b}^{f P}\right)^*[\bar{\n}_{\a}\g^{\m}L\n_{\b}][\bar{f}\g_{\m}Pf]
\eea
where $P \in {L,R}$, $L=1-\g^5$, $R=1+\g^5, $ $f=e, u, d$ and $\e_{\a \b}^{fP}$ is the deviation from standard interactions. 
There are model dependent bounds on these NSI parameters \cite{nsi0,sk}. In $R$-parity violating Supersymmetric models
these NSI parameters could be related to trilinear lepton number violating couplings \cite{adhi}. Also such parameters could be sizable
\cite{val} in unified supersymmetric models \cite{raby}. The model independent bounds have been discussed in \cite{nsi1}. The above NSI parameters can be reduced to the effective parameters as:
\be
\label{eq:nsieffec}
\e_{\a \b}=\sum_{f,P} \e_{\a \b}^{fP}\frac{n_f}{n_e} 
\ee
where $n_f$ is the number density of fermion, $n_e$ is the electron number density. In neutrino oscillation experiments this effective parameter ($\e_{\a\b}$) corresponds to the replacement in the matter interaction part of the evolution of flavoured neutrinos. This change can be seen as below:
\bea
\label{eq:hmatter}
H_{matter} = \pmatrix{1 & 0 & 0 \cr 0 & 0 & 0 \cr
 0
 & 0 & 0}\rightarrow \left[\pmatrix{1 & 0 & 0 \cr 0 & 0 & 0 \cr
 0
 & 0 & 0}+\pmatrix{\e_{ee} & \e_{e\mu} & \e_{e\tau} \cr \e_{e\mu}^* & \e_{\mu\mu} & \e_{\mu\tau} \cr
 \e_{e\tau}^*
 &\e_{\mu\tau}^* & \e_{\tau\tau}}\right]
\eea
In general, $\e_{e\mu}$, $\e_{e\tau}$ and $\e_{\mu\tau}$ could be complex. However, for our numerical analysis we have considered $\e_{e\mu}$ and  $\e_{e\tau}$ to be real. If we assumne uncorrelated errors,  the bounds on $\e_{\a\b}$ can be approximately written as \cite{nsi1}
\bea
\label{eq:nsibound}
\e_{\a\b}\lesssim \left[\sum_P((\e_{\a\b}^{ep})^2+(3\e_{\a\b}^{up})^2+(3\e_{\a\b}^{dp})^2)\right]^{1/2}
\eea
The NSIs $\e_{ee}$, $\e_{e\mu}$ and $\e_{e\tau}$ play significant role in $\nu_e \rightarrow \nu_\mu $ oscillation channel 
which we have considered. The bounds for these \cite{nsi1} in the context of 
neutrino oscillation for neutrinos passing through neutral earth like matter is $\e_{ee}<4$, $\e_{e\mu} < 0.33$ and $\e_{e\tau}<3$.

In vacuum, flavor eigenstates $\nu_\alpha$ may be related to  mass eigenstates of neutrinos $\nu_i$ as
\be
\vert\nu_\alpha>=\sum_{i}  U_{\alpha i}\vert\nu_i>
, \;\;U=R_{23} R_{13}(\delta) R_{12}
\;\;\quad \textrm{and}
\qquad i=1, 2, 3,
\ee
where $U$ is PMNS matrix \cite{pmns} and $R_{ij}$ are the rotation
matrices and $R_{13}(\delta ) $ contains the $CP$ violating phase $\delta$ signifying the complex rotation \cite{val}.
General probability expression for oscillation of neutrino of flavor $l$ to
neutrino flavor $m$ in matter (satisfying adiabatic condition for the density
of matter) is given by
\bea
\label{eq:genprob}
 P(\nu_{l}\rightarrow\nu_{m})=\delta_{lm}-4\sum_{i>j}  Re[J_{ij}^{lm}]
\sin^2\Delta^{'}_{ij}+2 \sum_{i>j} Im[J_{ij}^{lm}] \sin 2 \Delta^{'}_{ij}
\eea
where
\be
\label{eq:Jij}
J_{ij}^{lm}=U^{'}_{li} {U^{'\ast}_{lj}} {U^{'\ast}_{mi}} U^{'}_{mj} ,
\ee
\be
\label{eq:Dij}
\Delta^{'}_{ij}=\frac{\Delta ^{'}m_{ij}^2 L}{4E} . 
\ee 
Here 
\be
\label{eq:mij}
\Delta^{'} m_{ij}^2={m^{'}_{i}}^2-{m^{'}_{j}}^2 
\ee
and label ($\;^{'}\;$) indicates the
neutrino matter interaction induced quantities corresponding to those quantities
in vacuum.

We discuss in brief the perturbative approach for evaluating the induced quantities and for obtaining the probability of oscillation $\nu_e \rightarrow \nu_\mu$ for neutrinos passing through earth matter.
The diagonal neutrino mass matrix is approximately given by
\be
\label{eq:massneu}
m \approx \Delta m_{31}^2 diag(0,\a,1).
\ee
The effective Hamiltonian  induced by interaction of matter with neutrinos
is written in weak interaction basis as
\be
\label{eq:hamil}
 H \approx R_{23}MR_{23}^\dag
\ee
where
\be
 \label{eq:massmatter}
 M= \frac{\Delta m_{31}^2}{2E}R_{13}(\delta )R_{12}\; {m \over \Delta m_{31}^2}\;R_{12}^\dag R_{13}(\delta)^\dag
 +\frac{\Delta m_{31}^2}{2E} diag(A,0,0) +\frac{\Delta m_{31}^2}{2E} R_{23}^\dag\pmatrix{Z & X & Y \cr X^{\ast} & B & C \cr
 Y^{\ast}
 & C^{\ast} & D} R_{23}\; .
\ee
In equation (\ref{eq:massmatter})  
\bea
\label{eq:matternsi}
A=\frac{2E\sqrt{2}G_{F}n_{e}}{\Delta m_{31}^2},\;
 X=A \e_{e\mu}, \;Y=A\e_{e\tau}, \;
Z= A\e_{ee},\; B=A\e_{\mu\mu},\;
C=A\e_{\mu\tau},\;
D=A\e_{\tau\tau},
\eea
where $A$ is considered due to Standard model interaction of neutrinos with
electron 
and $G_{F}$ is the Fermi constant and $n_{e}$ is the electron number density written as
$n_e =(0.5 N_A) \rho $ and $N_A$ is Avogadro's number and $\rho$ is the matter density in gm/cc. $\e_{ee}$, $\e_{e\mu}$ , $\e_{e\tau}$, $\e_{\mu\mu}$, $\e_{\mu\tau}$
 and $\e_{\tau\tau}$  are considered due to NSI
 of neutrinos with matter. 
 We consider magnitude of these NSI parameters except $\e_{ee}$
 not higher than $\a$   in using perturbation method. As $\e_{ee}$ has been
 considered in the leading part of the Hamiltonian, our perturbative result is fine even for its highest experimentally allowed value. In our numerical analysis we have considered even the uppermost allowed values of other NSI parameters
 \cite{nsi1}. In equation (\ref{eq:massmatter}), ($\; ^{*} \;$)  is
 denoted for complex conjugation.

The mixing matrix $ U^{\prime}$ can be found out as $ U^{\prime}=
R_{23} \; W $. Here $W$
is the normalized eigenvectors of $\Delta m_{31}^2M /(2 E)$
calculated through perturbative technique similar to the one adopted  in \cite{perturb}.  We have  taken into account only the non-degenerate perturbative approach.
Let us consider the case where NSIs are present and where $ \sin \theta_{13} $ is 
small and of the order of $\alpha$ or less. 
M in equation (\ref{eq:massmatter}) can be written as $ M = M^{(0)}+M^{(1)}+M^{(2)} $ where $M^{i}$ contains terms
of the order of $\alpha^{i}$.
Then we can write
\bea
\label{eq:masspert}
 M^{(0)}=\frac{\Delta m_{31}^2}{2E} \; diag(A',0,1), \;\; M^{(1)}=
 \frac{\Delta m_{31}^2}{2E}\pmatrix{\alpha s_{12}^2 & b & a\cr b^{\ast}
 & \alpha c_{12}^2+ c_{23} c- s_{23} d & s_{23} c + c_{23} d \cr a^{\ast} & c_{23} e - s_{23} f & s_{23} e + c_{23} f}, \nonumber \\
 M^{(2)}=\frac{\Delta m_{31}^2}{2E}\pmatrix{s_{13}^2 & 0 & -e^{-i\delta}\alpha c_{13} s_{12}^2 s_{13} \cr 0 & 0 & -e^{-i\delta}\alpha c_{12} s_{12} s_{13} \cr -e^{i\delta}\alpha c_{13} s_{12}^2 s_{13} & -e^{i\delta}\alpha c_{12} s_{12} s_{13} & - s_{13}^2}
\eea
where
\bea
\label{eq:nsi}
 A'=A (1+\e_{ee}),\;  a &=&c_{23}Y+e^{-i\delta}s_{13} +Xs_{23} ,\;
b=c_{23}X+c_{12}\alpha s_{12} -Ys_{23} ,\nonumber \\
c&=&B c_{23}- C^* s_{23}, \;
d= C c_{23}- D s_{23}, \;
e= C^* c_{23} + B s_{23}, \;
f = D c_{23} + C s_{23} 
\eea
The  eigenvalues of $H$ upto second order in $ \alpha $ are
\bea
\label{eq:eigenvalues}
 {{m^{'}_1}^2 \over 2 E}
 &\approx &\frac{\Delta m_{31}^2}{2E}\left[A'+\alpha s_{12}^2+s_{13}^2  +
 \frac{|b|^2}{A'}+\frac{|a|^2}{(-1+A')}\right],  \nonumber \\
{ {m^{'}_2}^2 \over 2 E}
&\approx &\frac{\Delta m_{31}^2}{2E}\left[\alpha c_{12}^2-\frac{|b|^2}{A'} + c c_{23}^2-\left(d c_{23}^2 +  c s_{23}^2\right)^2
\right], \nonumber \\ 
 {{m^{'}_3}^2 \over 2 E}
&\approx &\frac{\Delta m_{31}^2}{2E}\left[1- s_{13}^2+\frac{|a^\ast|^2}{(1-A')} + f c_{23}^2 + e s_{23} + \left(e c_{23}-f s_{23}\right)^2\right]
\eea
Using the eigenvalues of H given in equation(\ref{eq:eigenvalues}) to equation(\ref{eq:mij}), the term $\Delta^{'}_{ij}$ in equation(\ref{eq:genprob}) can be calculated ($\Delta^{'}_{ij}$ is defined in equation(\ref{eq:Dij})). From equation (\ref{eq:eigenvalues}), it is seen that NSIs are present in the eigenvalues of H but those are of the order of $\alpha$ or $\alpha^2$. The oscillation probability $P(\nu_e \to \nu_\mu)$ is calculated upto order $\alpha^2$. As the non-zero terms in $J^{lm}_{ij}$ in equation(\ref{eq:genprob}) calculated using equation(\ref{eq:Jij}) is already at order $\alpha^2$, only terms which are zeroth order in $\alpha$ is considered in calculating $\Delta^{'}_{ij}$. Hence, the oscillation probability, $P(\nu_e \to \nu_\mu)$, in the presence of NSI upto order $\alpha^2$ can be written as:
\bea
\label{eq:probcomp}
 P(\nu_e \to \nu_\mu)
 \approx \frac{4c_{23}^2}{A'^2}|b|^2
 \sin^2\left( \frac{\Delta m_{31}^2A'L}{4E}\right) 
 +\frac{4s_{23}^2}{(1-A')^2}|a|^2
 \sin^2\left( \frac{\Delta m_{31}^2(1-A')L}{4E}\right)  \nonumber \\
 +\frac{8s_{23}c_{23}}{A'(1-A')}Re\left(ba^{*}\right)
 \sin\left( \frac{\Delta m_{31}^2A'L}{4E}\right)\cos\left( \frac{\Delta m_{31}^2L}{4E}\right)\sin\left( \frac{\Delta m_{31}^2(1-A')L}{4E}\right)  \nonumber \\
 +\frac{8s_{23}c_{23}}{A'(1-A')}Im\left(ba^{*}\right)
 \sin\left( \frac{\Delta m_{31}^2A'L}{4E}\right)\sin\left( \frac{\Delta m_{31}^2L}{4E}\right)\sin\left( \frac{\Delta m_{31}^2(1-A')L}{4E}\right) 
\eea 
 For NSI terms $X=Y=Z=0$ in  equation (\ref{eq:probcomp}),  the probability expression reduces to that for the standard model interaction of neutrinos with matter. 
 
\section{Bimagic conditions on neutrino energy} 
 If we want the probability, $P(\nu_e \to \nu_\mu)$ to be independent of the $CP$ violating phase $\delta$  and $\theta_{13}$ upto order $\a^2$ then we have to use the condition:
\be
\label{eq:cond}
\sin\frac{\Delta m_{31}^2(1-A')L}{4E} = 0
\ee
One may note here that this corresponds to two different conditions for two different hierarchies of neutrino masses. Considering $\e_{ee} =0$ in $A'$ and keeping $\e_{\e\mu}$ and $\e_{e\tau}$ less than $\a$ as required by the perturbation theory  one can see that this condition on neutrino energy is the same one as discussed earlier without NSIs \cite{dighe}. For a given length $L$ of the baseline above condition constrains the neutrino energy $E$ as
\be
\label{eq:magenergy}
 E=\Delta m_{31}^2/\left( \pm 4n\pi/L +2\sqrt{2}G_{F}n_{e}\; (1 + \e_{ee})\right)\; .
\ee 
As long as $\e_{ee}$ is unknown it seems that this magic energy $E$ cannot be known. However, what is important in our work
is to know the possible range of this magic energy depending on the presently allowed range of $\e_{ee}$ which is less than 4. 
Using eq.(\ref{eq:cond}) in eq.(\ref{eq:probcomp}) we get 
\be
\label{eq:probmag}
 P(\nu_e \to \nu_\mu)
 \approx \frac{4c_{23}^2}{A'^2}|c_{23}X+c_{12}\alpha s_{12}-Ys_{23}|^2
 \sin^2\left( \frac{\Delta m_{31}^2A'L}{4E}\right)\; .
\ee
which is independent of $\theta_{13}$ and $\delta $. 

\begin{figure}[h!]
\centering
\begin{tabular}{cc}
\epsfig{file=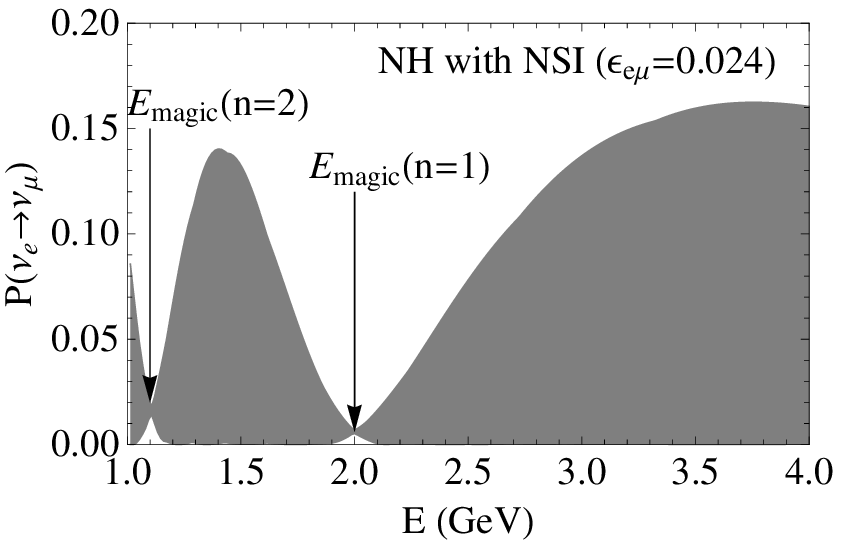,width=0.35\linewidth,clip=}&
\epsfig{file=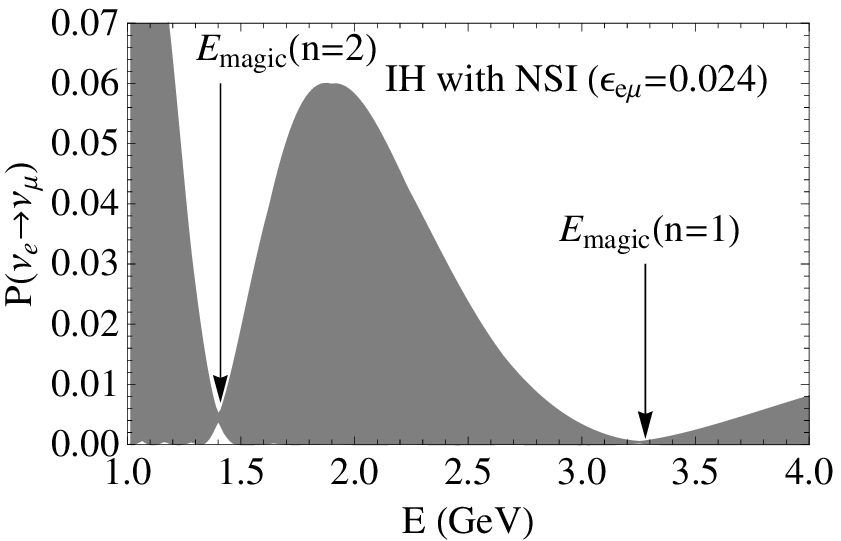,width=0.35\linewidth,clip=}\\
\epsfig{file=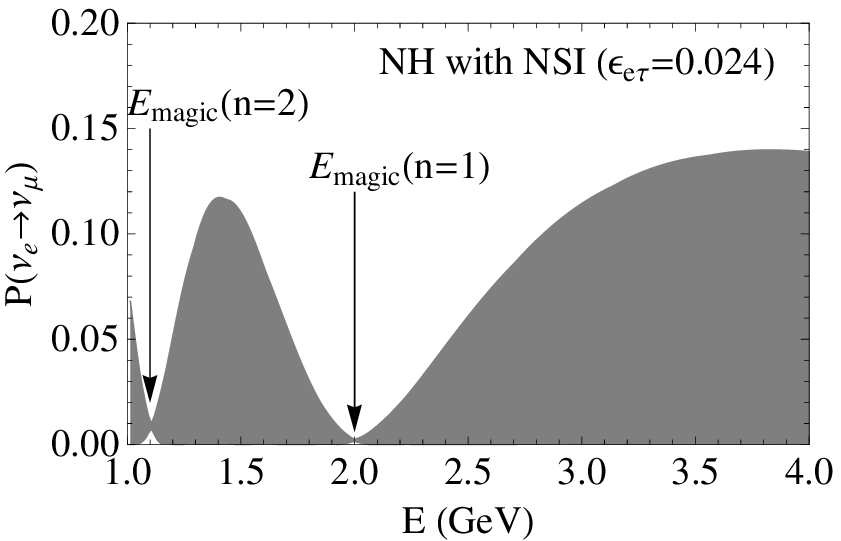,width=0.35\linewidth,clip=}&
\epsfig{file=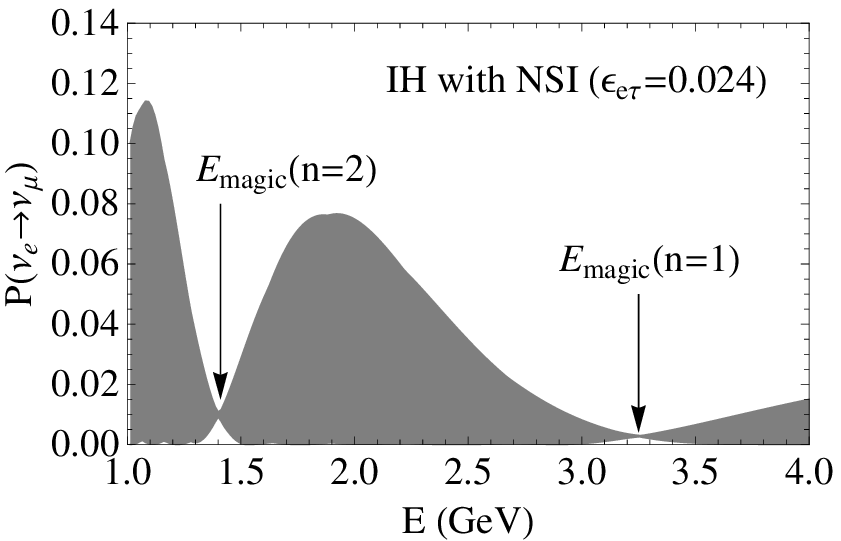,width=0.35\linewidth,clip=}
\end{tabular}
\caption[] {{\small Plots for probability of oscillation $\nu_e \rightarrow \nu_{\mu} $ versus neutrino energy ($E$) 
varying $\delta$ over the entire range of $0$ to $2 \pi $ and $\theta_{13} $ in the entire allowed range 0 to $12^{\circ}$ .}}
\label{fig:prob}
\end{figure}

In figure \ref{fig:prob} we have shown the probability of oscillation $\n_e \rightarrow \n_{\m}$ versus energy after varying $\theta_{13}$ in the entire allowed range of $0^{\circ}-12^{\circ}$  and also varying $\delta$ in the full range of $0-2 \pi $ for $L = 2540 $ Km for both normal (NH) and inverted hierarchies (IH). We have made the plots for two NSIs ($\e_{e\mu}$ and $\e_{e\tau}$) but considering one at a time. The value of the NSI considered is $\e_{e\mu}=0.024$ and $\e_{e\tau}=0.024$.  From the plots we can see that the probability becomes almost independent of $\delta$ and $\theta_{13}$ at the  magic energies. As for example, for $n=1$ in eq.(\ref{eq:magenergy}) these energies are ($E=1.9(3.3)$ GeV for NH(IH) respectively. Magic energies for
higher $n$ values gets smaller as seen from eq.(\ref{eq:magenergy}) and the figure.

Next we discuss the bimagic conditions on neutrino energy. As discussed in \cite{dighe} the sensitivity of the hierarchy is maximum if one of the hierarchies (say NH or IH) obey the
condition in eq.(\ref{eq:cond}) for which probability is independent of $\theta_{13}$  and $\delta $ whereas for the other hierarchy the probability has maximum dependence with $\theta_{13}$     
and $\delta $ which can be achieved by imposing the condition 
\be
\label{eq:cond2}
\sin\frac{\Delta m_{31}^2(1-A')L}{4E}=\pm 1  \; . 
\ee
These two conditions mentioned in (\ref{eq:cond}  ) and (\ref{eq:cond2}) can be rewritten in two different ways: 
(a) conditions for IH with $\delta $ and $\theta_{13}$ 
independence and NH with maximum dependence to those - which can be written as:   
\bea 
\label{eq:ihcond1}
\frac{|\Delta m_{31}^2|(1+|A'|)L}{4E\hbar c} = n\pi  \; \rm{for~IH}
\eea
\bea
\label{eq:ihcond2}
\frac{|\Delta m_{31}^2|(1-|A'|)L}{4E\hbar c} = (m-1/2)\pi \; \rm{for~NH}
\eea
where n, m are integers and $n>0$. 

(b) conditions for NH  with  $\delta $ and $\theta_{13}$ 
independence and IH with maximum dependence to those - which can be written as:
\bea
\label{eq:nhcond1}
\frac{|\Delta m_{31}^2|(1-|A'|)L}{4E\hbar c} = n' \pi  \;  \rm{for~NH}
\eea
\bea
\label{eq:nhcond2}
\frac{|\Delta m_{31}^2|(1+|A'|)L}{4E\hbar c} = (m'-1/2)\pi \; &\rm{for~IH}&
\eea
where $n'$ and $m'$ are integers and $n' \neq 0$ and $m' > 0$. 

Solving the equations (\ref{eq:ihcond1}) and (\ref{eq:ihcond2}) one gets the length
of the baseline ($L$) as
\bea
L \mbox{(Km )} =  \frac{(n-m+1/2) \pi \hbar  c}{ \sqrt{2} G_F n_e (1+\e_{ee})} \approx   
16260.5 \times  \frac{(n-m+1/2) }{ \rho (gm/cc) (1+\e_{ee})} \nonumber
\eea
which implies
\bea
\rho L \mbox {(Km gm/cc)} \approx 16260.5 \times (n-m+1/2)/(1+\e_{ee})
\eea
and for the inverted hierarchy the energy $E_{IH}$  with  $\delta $ and $\theta_{13}$ 
independence   as
\bea
E_{IH} \mbox{(GeV)} = \frac{1}{2\pi \hbar c\mbox{(GeV Km)}} \frac{|\Delta m_{31}^2| \mbox{(GeV}^2) L\mbox{(Km)}}{(n+m-1/2)}
\eea
 
Similarly, solving the equations  (\ref{eq:nhcond1}) and (\ref{eq:nhcond2}) one gets the length
of the baseline ($L'$) as
\bea
\rho L' \mbox{(Km gm/cc)} \approx  16260.5 \times (m'-n'-1/2)/(1+\e_{ee})  
\eea
and for the normal hierarchy the energy $E_{NH}$   with  $\delta $ and $\theta_{13}$ 
independence   as
\bea
E_{NH} \mbox{(GeV)} = \frac{1}{2\pi \hbar c\mbox{(GeV Km)}}  \frac{|\Delta m_{31}^2| \mbox{(GeV}^2) L'\mbox{(Km)}}{(n'+m'-1/2)}
\eea
Firstly, for $\e_{ee}=0$ one can get one possible solution for common baseline i.e., $L = L'$ to be about 2540 Km
if the choices are made as follows:
$n=1$, $m=1$ and $n'=1$ and $m'=2 $ .  The neutrino energy  $E_{IH} \approx 3.3$ GeV and $E_{NH} \approx 1.9$ Gev. However,
one could get more common solutions for bimagic baseline with $L=L'$ for length by considering suitable choices of $m,n,m' $ and $n'$ for which $n-m=m'-n'-1$;  $n-m$ could be 1 or 4 resulting in $L=L'> 6000$ Km. As for example, 
considering $\rho \approx 4$ gm/cc with $n-m=1$ the length is about $6100 $ Km.

As the present upper bound of some of the NSIs could be quite large $\gsim \a$ \cite{nsi1} which is not considered in our perturbative approach it is natural to ask what happen to such magic energies in the same 2540 Km baseline in those cases. We have checked numerically that even for highest allowed values of NSIs as for example,for $\e_{e\mu} = 0.33$ the $E_{NH}(E_{IH}) \approx 2.03(3.26)$ GeV; for $\e_{ee} = 4.0$ the $E_{NH}(E_{IH}) \approx 1.18(3.18)$ GeV.  However, for $\e_{e\tau} \gsim 0.5$   it is difficult to get bimagic energies although we have 
presented numerical analysis of discovery limits of various oscillation parameters for that also. For $\e_{e\tau} =0.5$
the bimagic energies are the $E_{NH}(E_{IH}) \approx 2.02(3.57)$ GeV. 
It is important to note that all these bimagic
 energies are within 1-5 GeV which is the full neutrino energy range in our analysis. Interestingly, one can check here that the perturbative results for
 bimagic baseline length and the energies hold good even for higher values of $\e_{ee} >> \a $. Even for such large $\e_{ee}$
 it is possible to obtain bimagic energies within 1-5 GeV for the same baseline length of 2540 Km. As for example, for $\e_{ee} =4$ considering $n=3$, $m=1$, $m'=4$ and $n'=1$ gives $L=L'\approx 2540$ Km and
 $E_{IH}$ and $E_{NH}$ obtained from perturbative approach are very near to the numerical values for bimagic energies mentioned above. 
As the bimagic energies are within 1-5 GeV, by considering this as the neutrino energy range in our numerical analysis, it may be expected to get  better discovery limits to hierarchy, $\theta_{13}$ and $CP$ violation for various choices of  NSIs. 

Now, we can write down the probabilities at the particular energies $E_{IH}$ and $E_{NH}$ according to the conditions discussed above. 
At $E \approx E_{IH}$ 
for condition satisfying  eq. (\ref{eq:ihcond1}) the probability is given by:
\bea
\label{probIH1}
P(\nu_e \rightarrow \nu_\mu)(IH) \approx \frac{4c_{23}^2}{|A'|^2}|b|^2
 \sin^2\left(\frac{\pi}{2}\left(n+m-\frac{1}{2}\right)\right)
\eea
and for the condition satisfying  eq. (\ref{eq:ihcond2}) the probability is given by:
\bea
\label{probIH2}
&&P(\nu_e \rightarrow \nu_\mu)(NH) \approx \frac{4c_{23}^2}{|A'|^2}|b|^2
 \cos^2\left( \frac{\pi}{2}\left(n+m-\frac{1}{2}\right) \right) 
 +\frac{4s_{23}^2}{(1-|A'|)^2}|a|^2
  \nonumber \\
 &&-\frac{8s_{23}c_{23}}{|A'|(1-|A'|)}\left[Re\left(ba^{*}\right)\cos^2\left(\frac{\pi}{2}\left(n+m-\frac{1}{2}\right)\right) 
 +\frac{1}{2}Im\left(ba^{*}\right)
 \sin\left( \pi\left(n+m-\frac{1}{2}\right)\right)\right]\eea
 At $E \approx E_{NH}$ 
for condition satisfying  eq. (\ref{eq:nhcond1}) the probability is given by
\bea
\label{probNH1}
P(\nu_e \rightarrow \nu_\mu)(NH) \approx \frac{4c_{23}^2}{|A'|^2}|b|^2
 \sin^2\left(\frac{\pi}{2}\left(n'+m'-\frac{1}{2}\right)\right) 
\eea
and for the condition satisfying  eq. (\ref{eq:nhcond2}) the probability is given by

\bea
\label{probNH2}
&&P(\nu_e \rightarrow \nu_\mu)(IH) \approx \frac{4c_{23}^2}{|A'|^2}|b|^2
\cos^2\left(\frac{\pi}{2}\left(n'+m'-\frac{1}{2}\right) \right) 
 + \frac{4s_{23}^2}{(1+|A'|)^2}|a|^2
   \nonumber \\
 &&+ \frac{8s_{23}c_{23}}{|A'|(1+|A'|)}\left[Re\left(ba^{*}\right)\cos^2\left(\frac{\pi}{2}\left(n'+m'-\frac{1}{2}\right)\right)
  - \frac{1}{2}Im\left(ba^{*}\right)\sin\left(\pi\left(n'+m'-\frac{1}{2}\right)\right)\right]\eea

One can see that at $E_{IH}=3.3$ GeV from eqs. (\ref{probIH1}) and (\ref{probIH2}) in the case of $\e_{e\tau}=\e_{e\mu}=0$
case i.e., $X=Y=0$ if $\theta_{13}$ also vanishes then $a=0$ and there is no difference
in $P_{IH}$ and $P_{NH}$. Same thing happens at $E_{NH}=1.9 $ GeV as seen in eqs. (\ref{probNH1}) and (\ref{probNH2}). This means that for $\e_{e\tau}=\e_{e\mu}=0$ case it is not so likely to get the hierarchy discovery limit
at $\theta_{13}=0$. However, on the contrary in presence
of these NSIs there is difference in $P_{IH}$ and $P_{NH}$ even for $\theta_{13}=0$.
So in presence of NSIs like $\e_{e\tau}$ and $\e_{e\mu}$ one could get discovery limit of hierarchy even at $\theta_{13}=0$.  However, in case of $\e_{ee}$ if $\theta_{13}=0$
then there is no difference between $P_{IH}$ and $P_{NH}$ and so it is not likely to get discovery limit at $\theta_{13}=0$.
These features are found in our numerical analysis.

In probing other NSIs like $\e_{\mu\mu}$, $\e_{\mu\tau}$ and $\e_{\tau\tau}$, oscillation channel $\nu_e\rightarrow \nu_{\mu}$  is not appropriate. This follows from the probability in equation (\ref{eq:probcomp}). For those NSIs considering the disappearance channel $({\nu_{\mu}\rightarrow \nu_{\mu}})$
 is appropriate. One cannot get any condition on neutrino energy in general to remove the dependence on $\delta$
 in the oscillation probability for this channel. However, it is found that upto the order $\alpha$ without imposing any condition on neutrino energy this oscillation probability is already independent of $\delta$ as shown 
below \cite{survprob}. 
\bea
\label{eq:survprob}
P(\nu_{\mu} \rightarrow\nu_{\mu})&=& 1 - 4 c_{23}^2 s_{23}^2 \sin^2 \frac{\Delta  m_{31}^2 L }{4 E} + 
4 c_{12}^2 c_{23}^2 s_{23}^2 \a \frac{\Delta m_{31}^2 L} {4 E} \sin \frac{\Delta m_{31}^2 L} {2 E} 
\nonumber \\ &+& 2 c_{23}^2 s_{23}^2 \left[ \left( c_{23}^2 -s_{23}^2 \right) \left(\e_{\m\m}-\e_{\tau\tau}\right)
- 4 c_{23} s_{23} Re(\e_{\mu\tau}) \right] \frac{\Delta m_{31}^2 A' L}{2 E} \sin \frac{\Delta m_{31}^2 L} {2 E} 
\nonumber \\ &-& 8 c_{23} s_{23} \left( c_{23}^2 -s_{23}^2 \right) \left[ c_{23} s_{23} \left(\e_{\m\m}-\e_{\tau\tau}\right) +
\left( c_{23}^2 -s_{23}^2  \right) Re(\e_{\mu\tau})   \right] A' \sin^2 \frac{\Delta  m_{31}^2 L }{4 E}
\eea 
To get the sensitivity of NSIs like  $\e_{\mu\mu}$, $\e_{\mu\tau}$ and $\e_{\tau\tau}$, this disappearance channel is appropriate but it is not much suitable for finding sensitivity to $\delta$ or discovery limits for $CP$ violation. As the probability upto order $\a$ is already independent of $\delta $ here one does not require the  bimagic conditions and as such there is no restriction on the length of the baseline and neutrino energy. We have not done this analysis separately but sensitivity of some of the above-mentioned  NSI parameters have been discussed in the  disappearance channel $(\nu_{\mu}\rightarrow \nu_{\mu})$ in 
\cite{nsi3}.

\section{Numerical Simulation} 
  As an outcome of the bimagic energy conditions in presence of NSIs the length of
  the baseline is $\approx 2540 $ Km.  To study the oscillation of ${\nu_e \rightarrow 
\nu_\mu} $ we have considered the experimental set-up and the detector characteristics as discussed in \cite{dighe} for a running time of 2.5 year.  We consider the neutrino factory having $ 5  \times 10^{21} $ muon decays per year with parent muon of energy 5 GeV and the magnetized totally active scintillator detector of 25 kt mass  with energy threshold of 1 GeV. The numerical simulation has been done by using GLoBES \cite{globes1}. In presenting the discovery limits of hierarchy,  $\sin^2 \theta_{13}$ and  $CP$ violation  in bimagic baseline, highest possible values for the NSIs have been considered. However, when we observe that the discovery limits are either covering the 
entire allowed region or not at all obtainable then we refrain from presenting those figures and instead we present discovery limits for some  lower values of NSIs.

\begin{figure}[H]
\centering
\begin{tabular}{ccc}
\epsfig{file=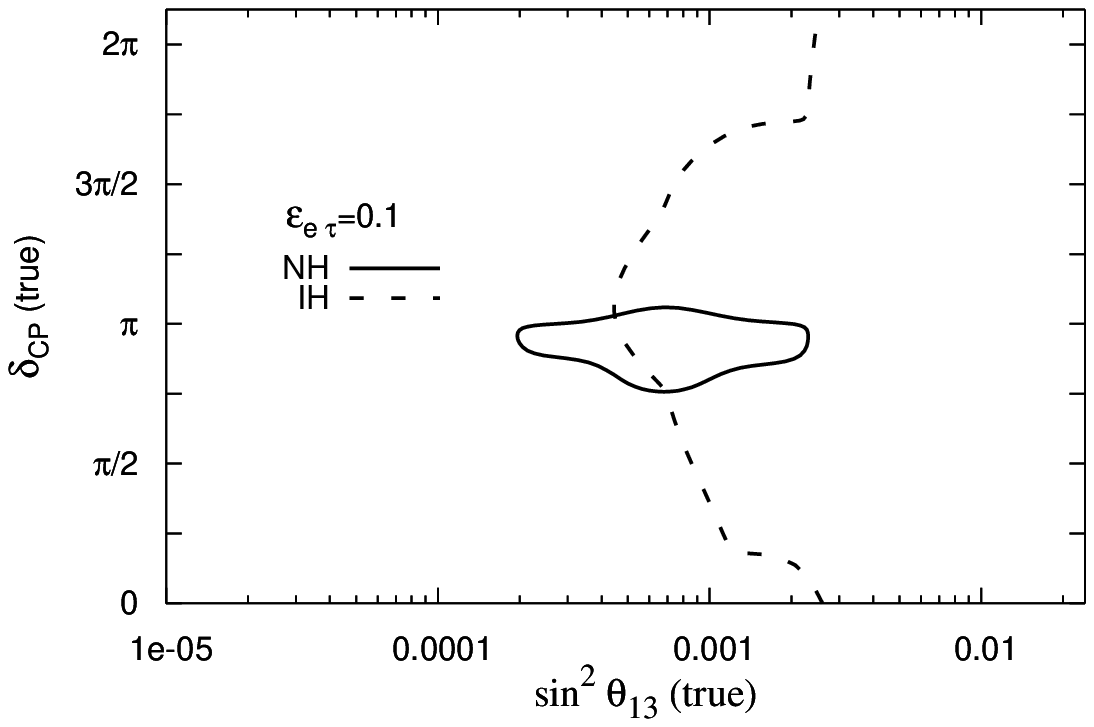,width=0.33\linewidth,clip=}&
\epsfig{file=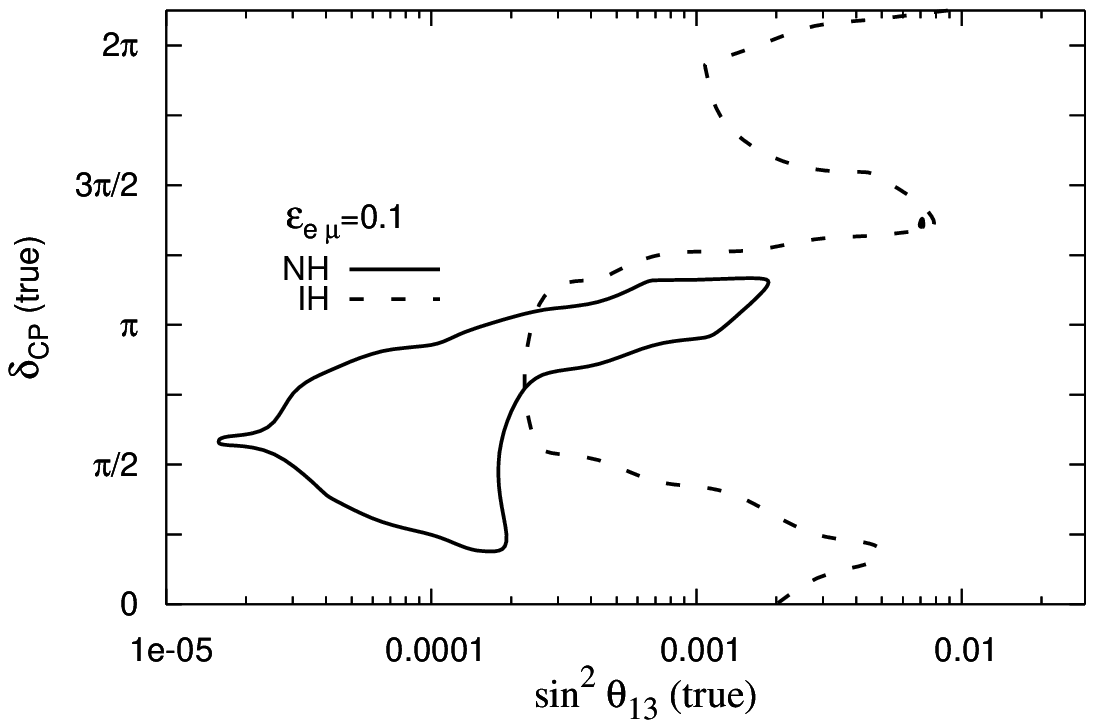,width=0.33\linewidth,clip=}& 
\epsfig{file=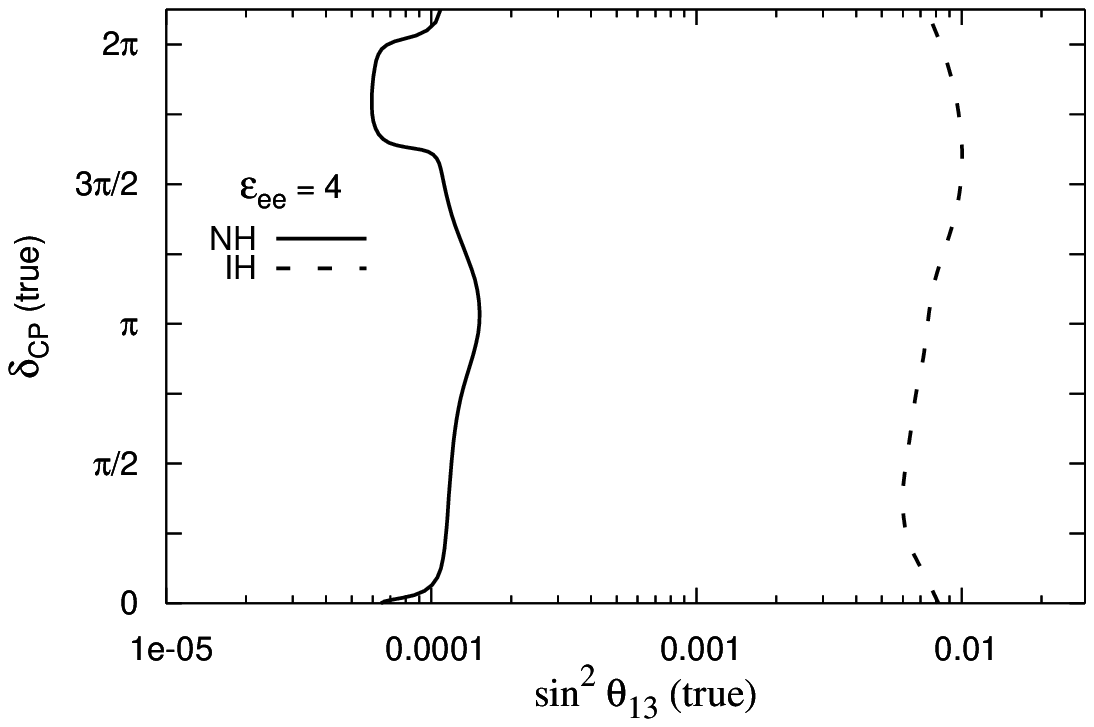,width=0.33\linewidth,clip=} 
\end{tabular}
\caption[] {{\small The 3$\sigma$ contours showing discovery limit of hierarchy for different NSIs  $\e_{e\tau}$, $\e_{e\mu}$ and $\e_{ee}$.}}
\label{fig:hierarchy}
\end{figure}

In figure \ref{fig:hierarchy} we have shown  for what value of $\delta $ and $\theta_{13}$ at 3 $\sigma $ confidence level one can identify the specific hierarchy which could be either normal or inverted. For $\epsilon_{e\tau}=3$ and $\epsilon_{e\mu}=0.33$ (one at a time)  NH could be discovered at any value of $\theta_{13}$ irrespective of any specific value of $\delta$ whereas for IH nowhere it is found to be discovered. So in our figures we have chosen some lower values of these two NSIs. Considering  $\epsilon_{e\mu}=0.1$ and $\epsilon_{e\tau}=0.1$ (one at a time) we have shown in
figure   \ref{fig:hierarchy} the discovery limit of hierarchy. From these figures  for favorable values 
of $\delta$ one could identify the  inverted hierarchy of nature  for $\epsilon_{e\tau}$ at  $\sin^2 \theta_{13}$ as small as $4 \times 10^{-4}$ and for $\epsilon_{e\mu}$ at $\sin^2 \theta_{13}$ as small as about $2.5 \times 10^{-4}$ . For normal hierarchy
however, it is found from the figures that  for $\e_{e\tau} =0.1$, only for $\delta$ in the range of 
$3 \pi/4$ to $ 5 \pi/4$ and $\sin^2 \theta_{13} \gsim 5 \times 10^{-4}$ one could reach the discovery limit. For other values of $\delta$ normal hierarchy can be identified for any value of $\theta_{13}$ including the zero value. Similarly, for $\e_{e\mu} =0.1$, only for $\delta$ in the range of about
$ \pi/4$ to $ 5 \pi/4$ and $\sin^2 \theta_{13} \gsim 2 \times 10^{-4}$ one could reach the discovery limit. Here also for other values of $\delta$ normal hierarchy can be identified for any values of $\theta_{13} $ including the zero value. For 
$\e_{ee} =4$ the normal hierarchy can be identified at $\sin^2\theta_{13} \gsim 10^{-4}$ and the inverted hierachy can be identified at $\sin^2\theta_{13}\gsim 10^{-2}$. 

\begin{figure}[H]
\centering
\begin{tabular}{ccc}
\epsfig{file=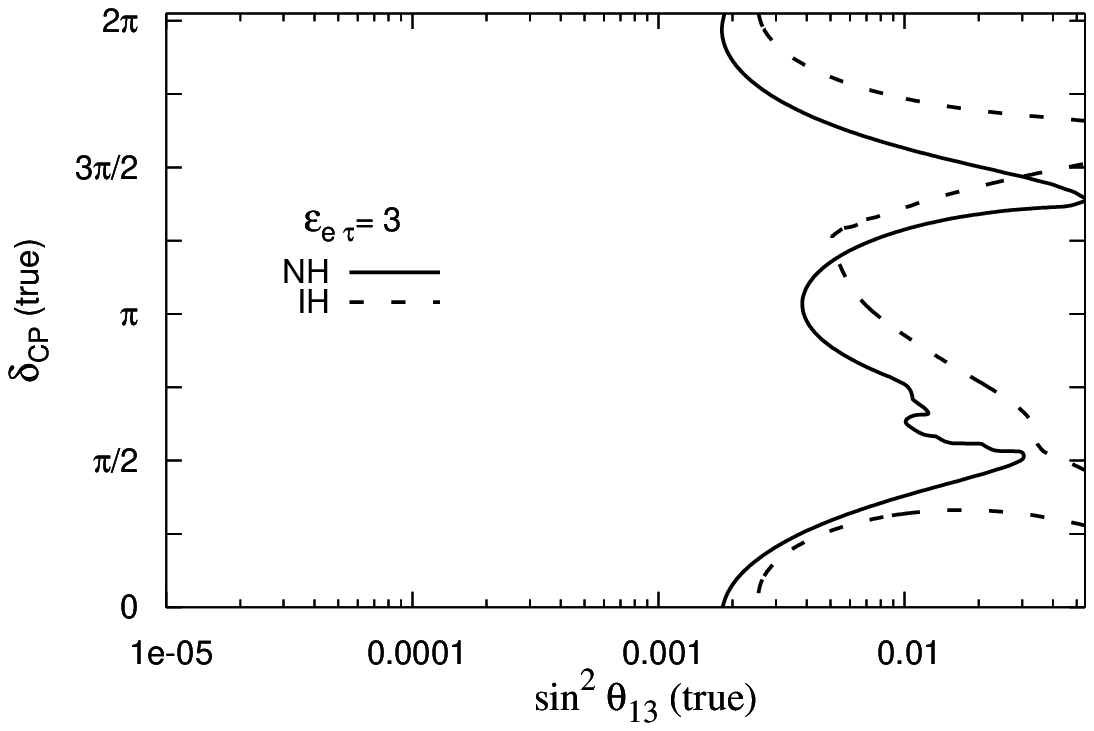,width=0.33\linewidth,clip=}&
\epsfig{file=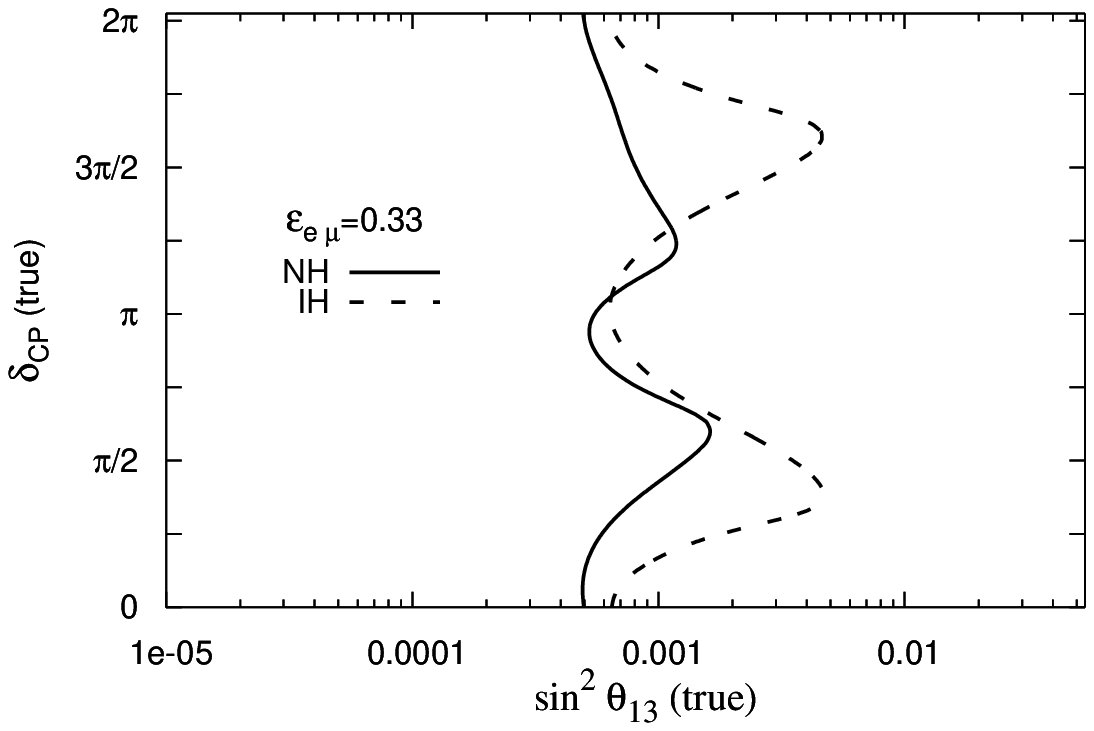,width=0.33\linewidth,clip=} &
\epsfig{file=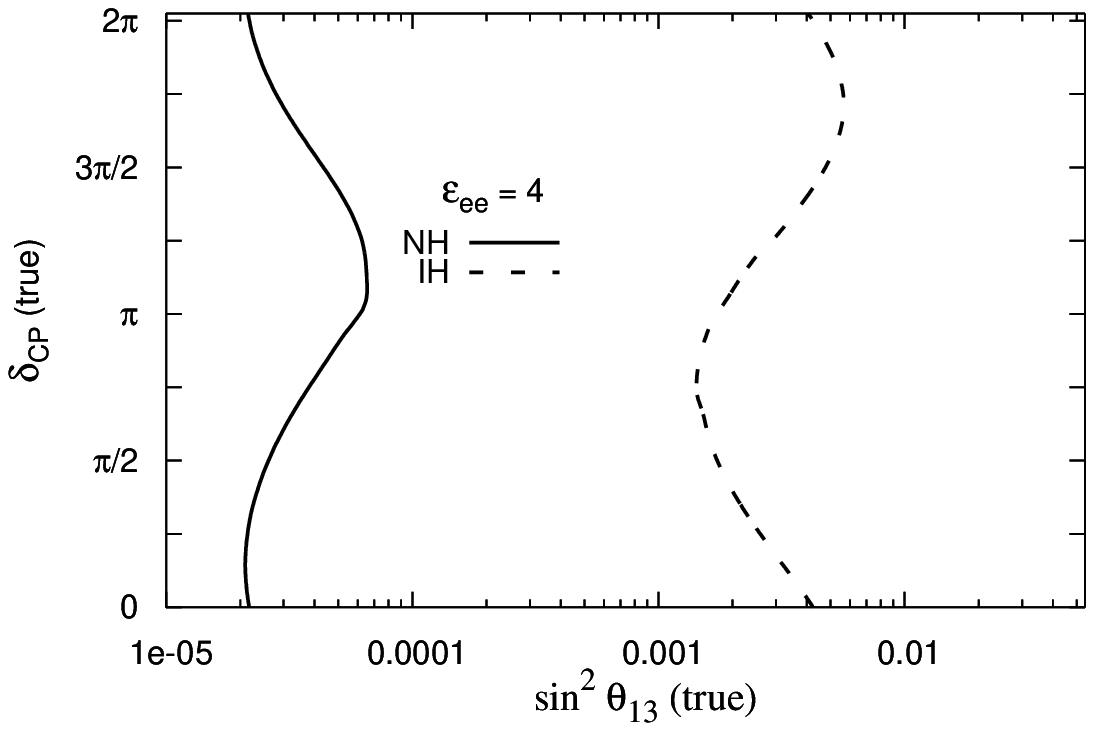,width=0.33\linewidth,clip=}
\end{tabular}
\caption[] {{\small The 3$\sigma$ contours showing discovery limits of $\theta_{13}$ for different NSI's $\e_{e\tau}$, $\e_{e\mu}$ and $\e_{ee}$.}}
\label{fig:theta13}
\end{figure}

\begin{figure}[H]
\centering
\begin{tabular}{ccc}
\epsfig{file=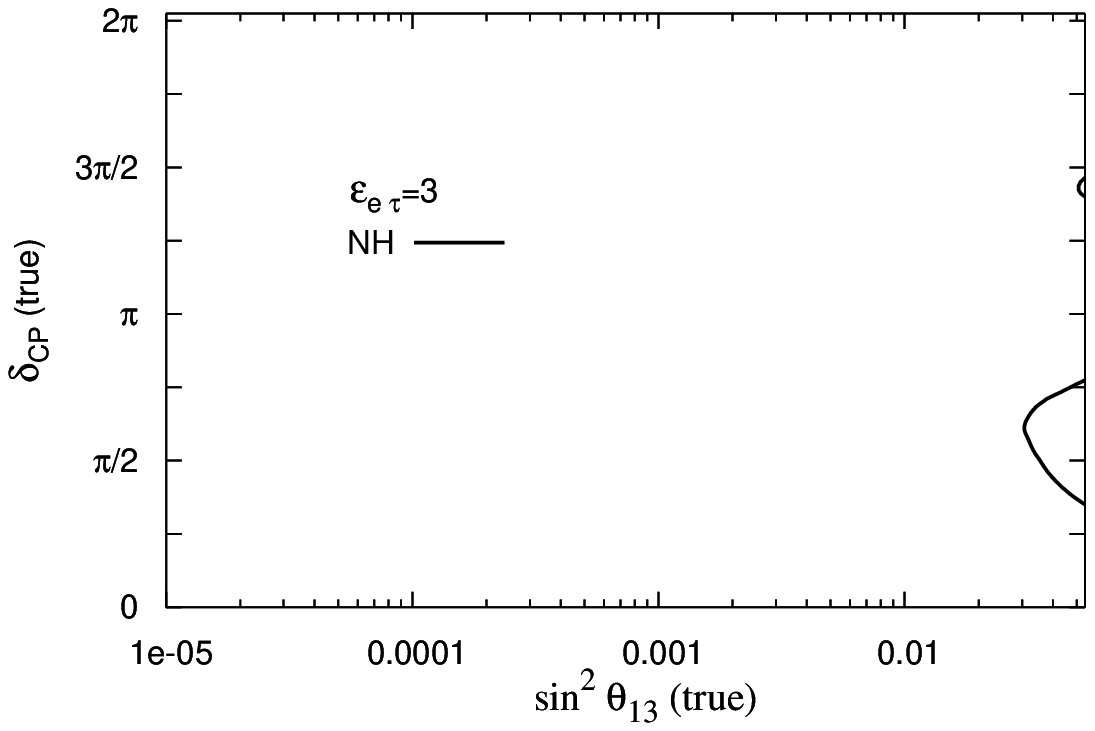,width=0.33\linewidth,clip=}&
\epsfig{file=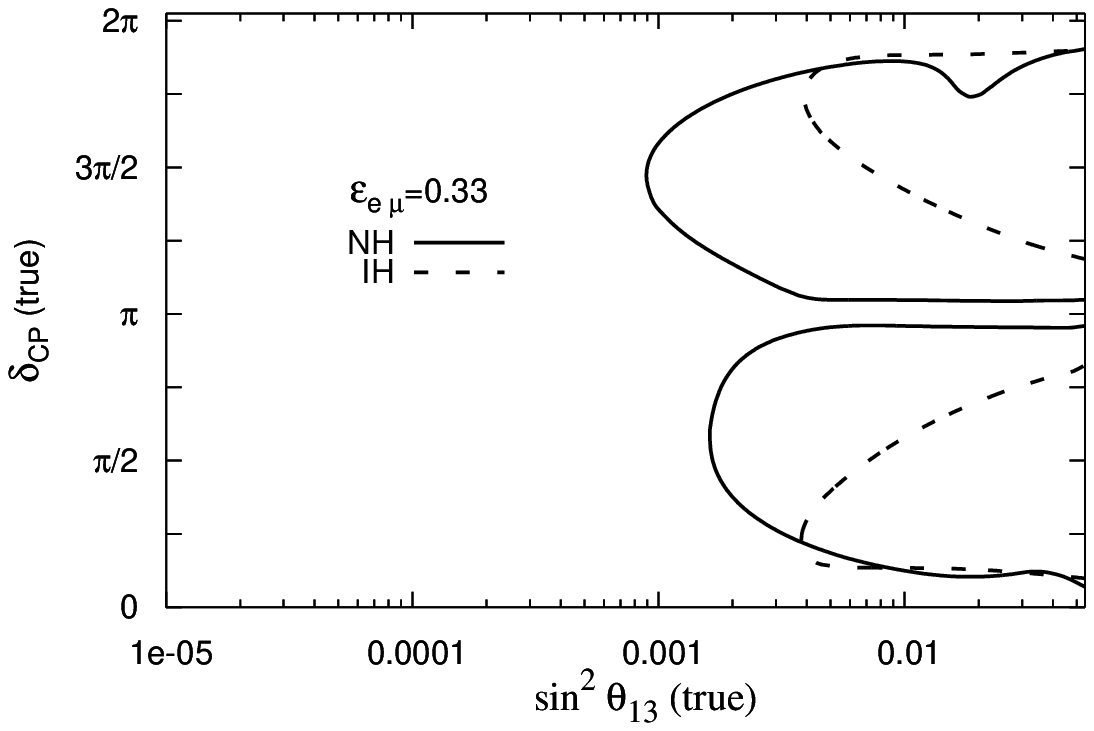,width=0.33\linewidth,clip=}&
\epsfig{file=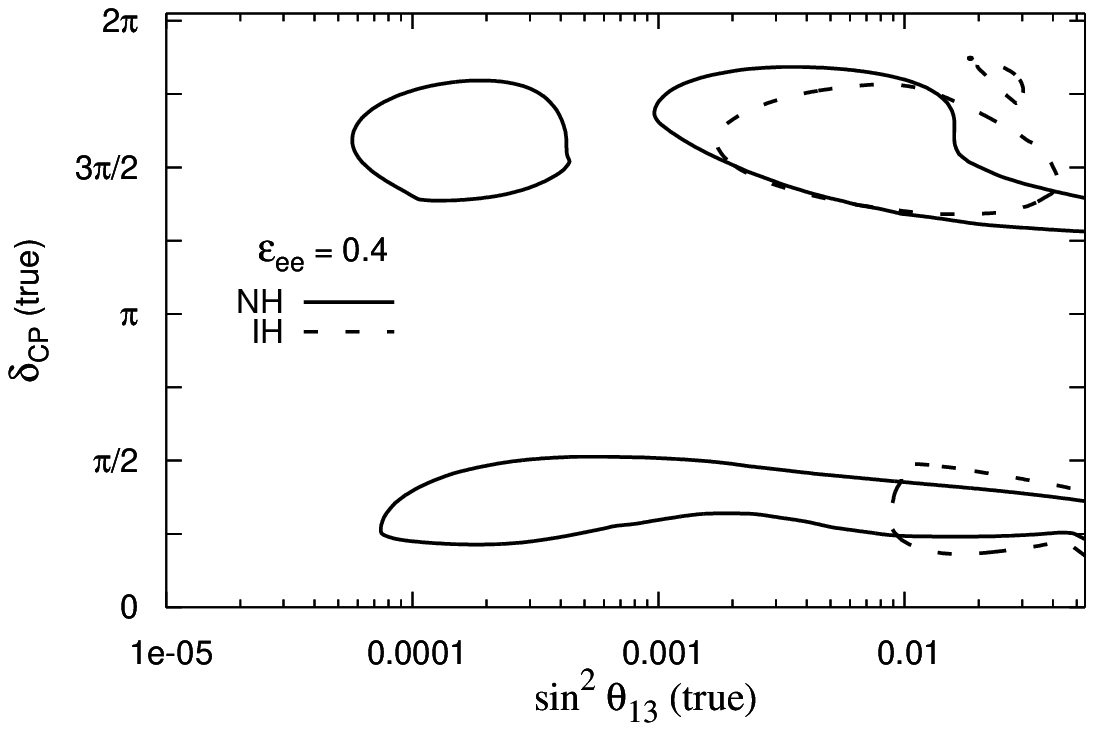,width=0.33\linewidth,clip=}
\end{tabular}
\caption[] {{\small The 3$\sigma$ discovery limits for the $CP$ violating phase $\delta$ for different NSI's $\e_{e\tau}$, $\e_{e\mu}$ and $\e_{ee}$.}}
\label{fig:delta}
\end{figure}
We have shown the discovery limits of $\theta_{13}$ in figure \ref{fig:theta13} at 3$\sigma$ confidence level. From figure \ref{fig:theta13} it can be seen that for NSI $\epsilon_{e \mu} \sim 0.33$  the discovery limits for $\sin^2\theta_{13}$ could be as low as  $\sin^2 \theta_{13} \sim 5.0 \times 10^{-4}$  and for $\epsilon_{e \tau} \sim 3.0$ the limit could be as low as $2.0\times 10^{-3}$ for both the hierarchies. However, for $\e_{ee} \sim 4$ this limit improves for NH and can be as low as $\sin^2 \theta_{13} \sim 2 \times 10^{-5}$, but for IH it could be as low as $\sin^2 \theta_{13} \sim 1.5 \times 10^{-3}$.  In figure \ref{fig:delta} we have shown the discovery limit of $CP$ violation for different NSI at $3 \sigma $ confidence level. The discovery limit for $CP$ violating region  is possible for $\epsilon_{e\tau} = 3.0$ at
$\sin^2 \theta_{13} \geq 2.5 \times 10^{-2}$ for $3 \pi/8 \leq \delta  \leq 3 \pi/4$  for NH. But for IH it is very difficult to obtain any discovery limit. However, for lower values of $\e_{e\tau}$ discovery limits could be easily obtained. 
For $\epsilon_{e\mu} = 0.33$  the discovery limit of $CP$ violating region  for NH is found  for $\sin^2 \theta_{13} \gsim  2 \times 10^{-3}$ with $ \pi/8 \lsim \delta \lsim 7 \pi/8$ and also  for $\sin^2 \theta_{13} \gsim  8 \times 10^{-4}$ with $ 9 \pi/8 \lsim \delta \lsim 15 \pi/8$. In the case of IH for the same value of $\e_{e\mu}$ the discovery limits are found for $\sin^2 \theta_{13} \gsim  4 \times 10^{-3}$ with $ \pi/8 \lsim \delta \lsim 3 \pi/4$ and also  for $\sin^2 \theta_{13} \gsim  4 \times 10^{-3}$ with $ 5 \pi/4 \lsim \delta \lsim 15 \pi/8$. For higher value of $\e_{ee}=4$ it is not possible to get discovery limit for $CP$ violating region. For $\epsilon_{ee} = 0.4$  the discovery limit of $CP$ violating region  for NH is found  for $\sin^2 \theta_{13} \gsim  7 \times 10^{-5}$ with $ \pi/4 \lsim \delta \lsim \pi/2$ and also  for $\sin^2 \theta_{13} \gsim  6 \times 10^{-5}$ with $ 5 \pi/4 \lsim \delta \lsim 7 \pi/4$. In the case of IH for the same value of $\e_{ee}$ the discovery limit are found for $\sin^2 \theta_{13} \gsim  9 \times 10^{-3}$ with $ \pi/8 \lsim \delta \lsim \pi/2$ and also  for $\sin^2 \theta_{13} \gsim  2 \times 10^{-3}$ with $ 5 \pi/4 \lsim \delta \lsim 7 \pi/4$.



\begin{figure}[h!]
\centering
\begin{tabular}{cc}
\epsfig{file=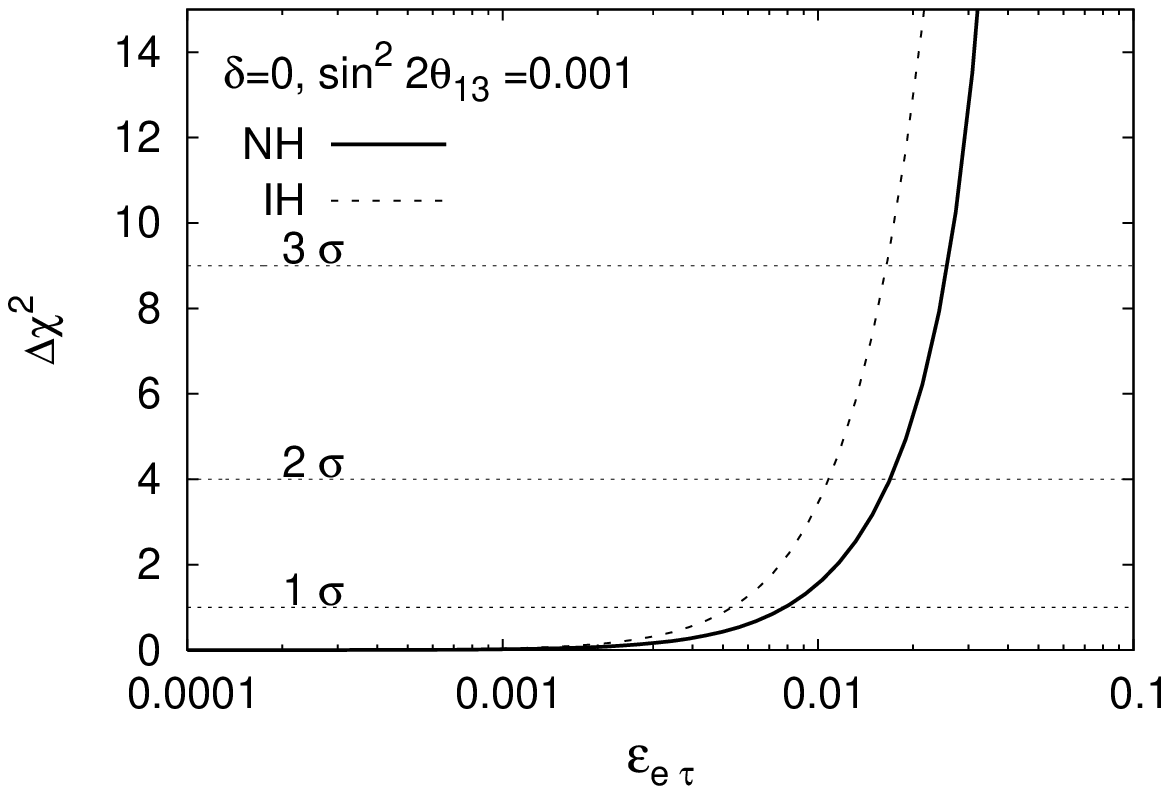,width=0.35\linewidth,clip=}&
\epsfig{file=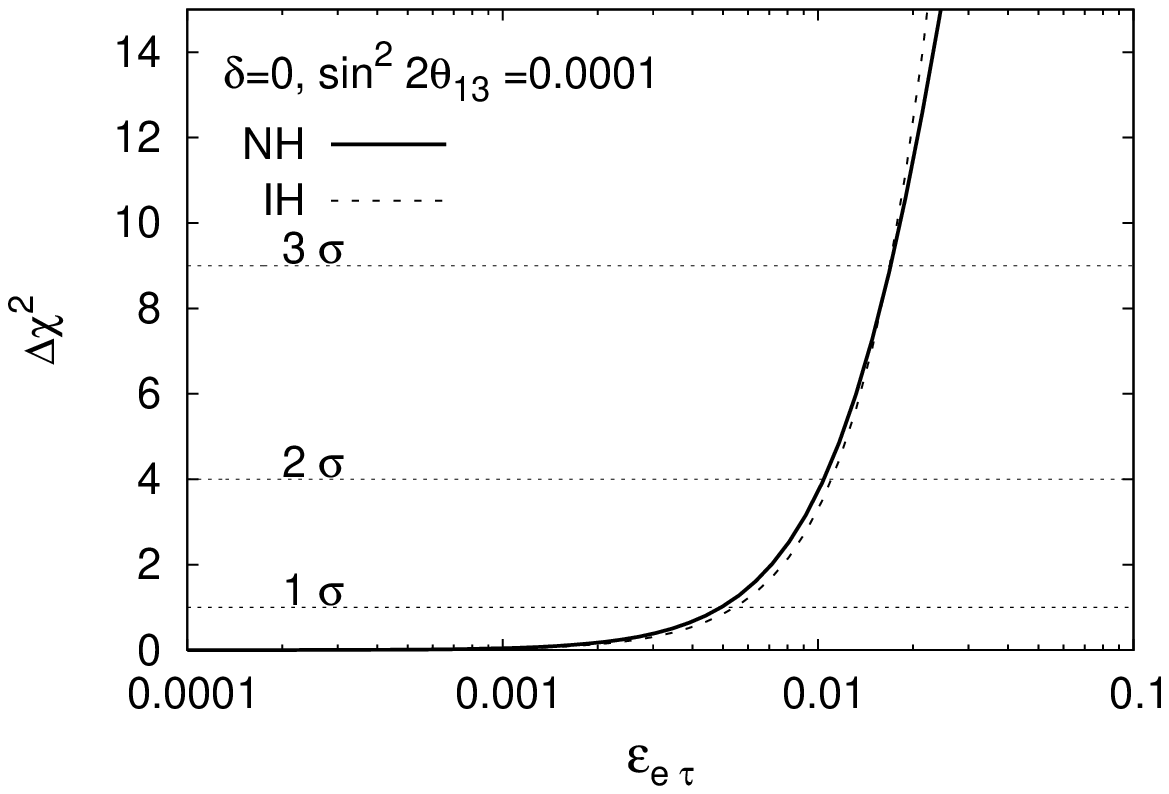,width=0.35\linewidth,clip=}\\
\epsfig{file=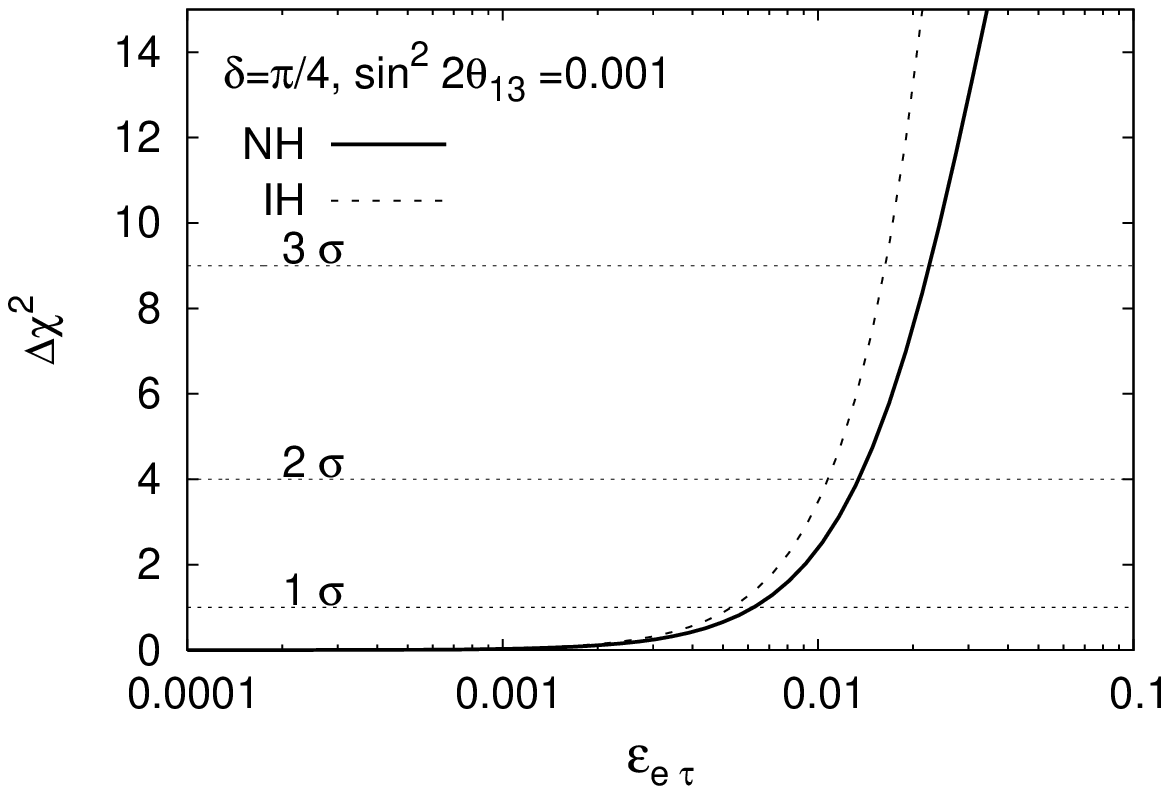,width=0.35\linewidth,clip=}&
\epsfig{file=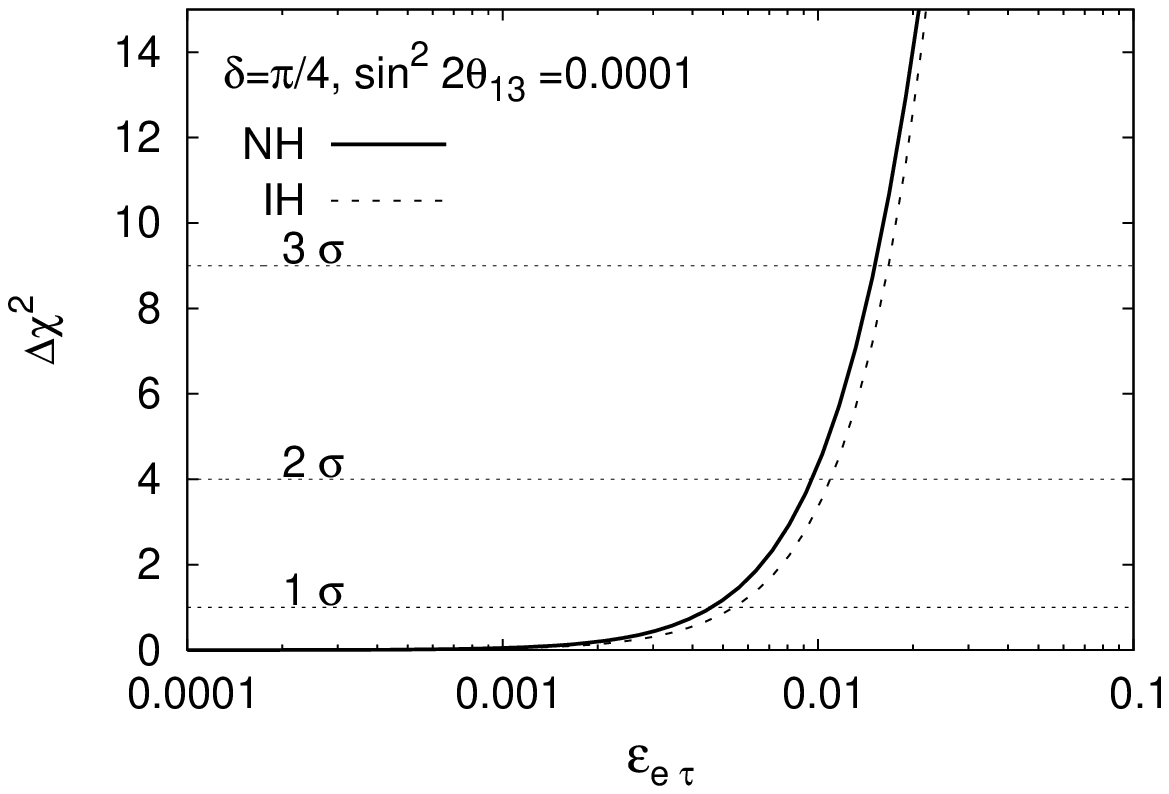,width=0.35\linewidth,clip=}
\end{tabular}
\caption[] {{\small Discovery limits of NSI ($\e_{e\tau}$) for different fixed values of $\theta_{13}$ and $\delta$ considering muon energy 5 GeV.}}
\label{fig:etnsifix}
\end{figure}

\begin{figure}[h!]
\centering
\begin{tabular}{cc}
\epsfig{file=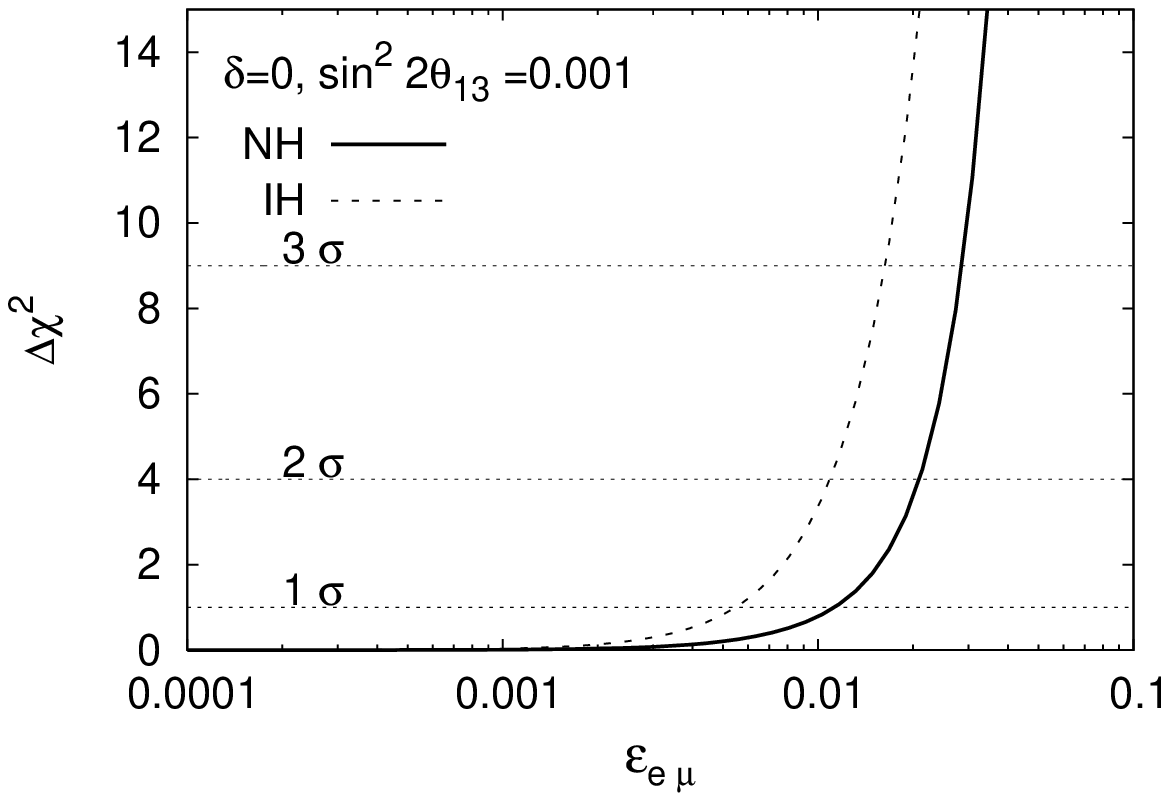,width=0.35\linewidth,clip=}&
\epsfig{file=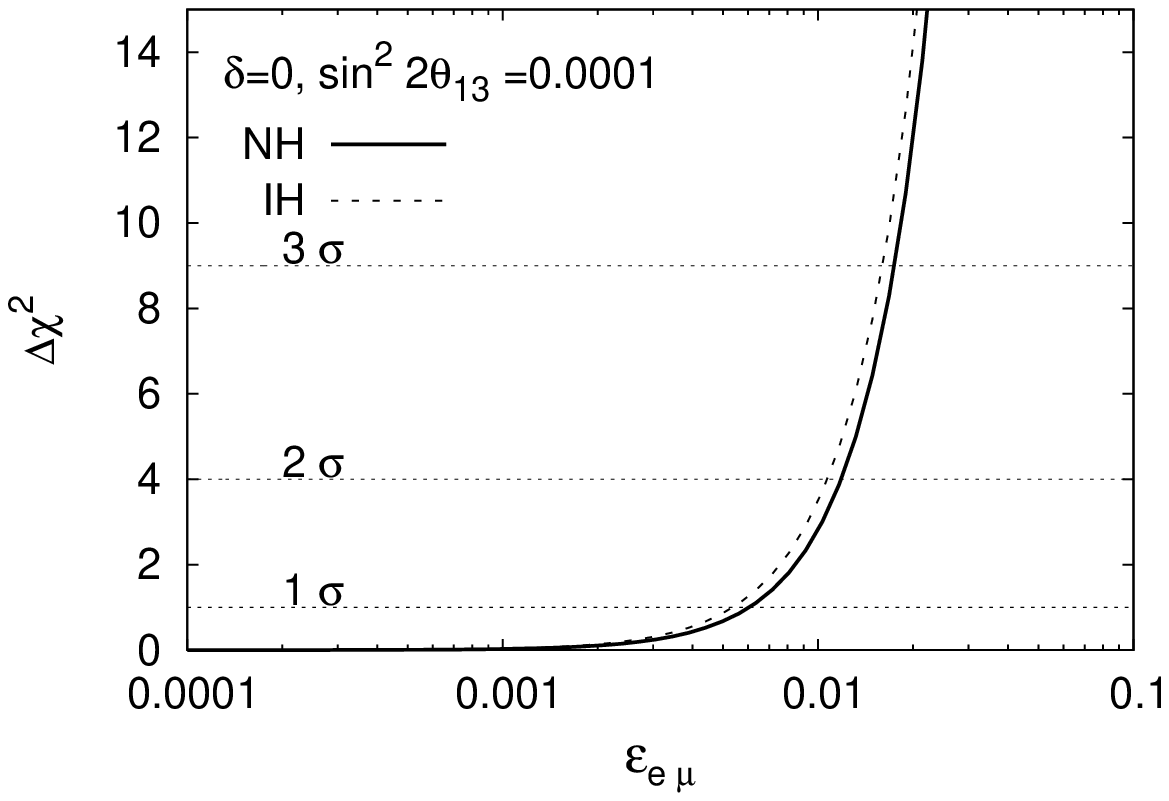,width=0.35\linewidth,clip=}\\
\epsfig{file=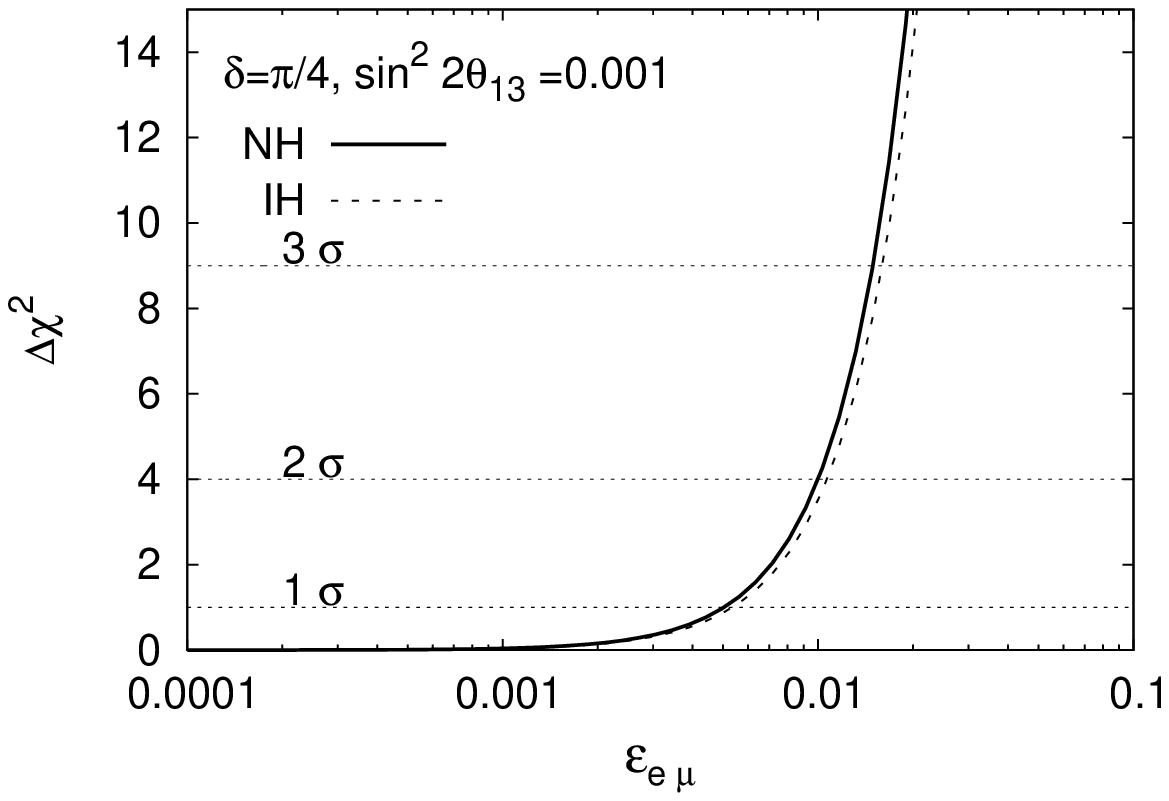,width=0.35\linewidth,clip=}&
\epsfig{file=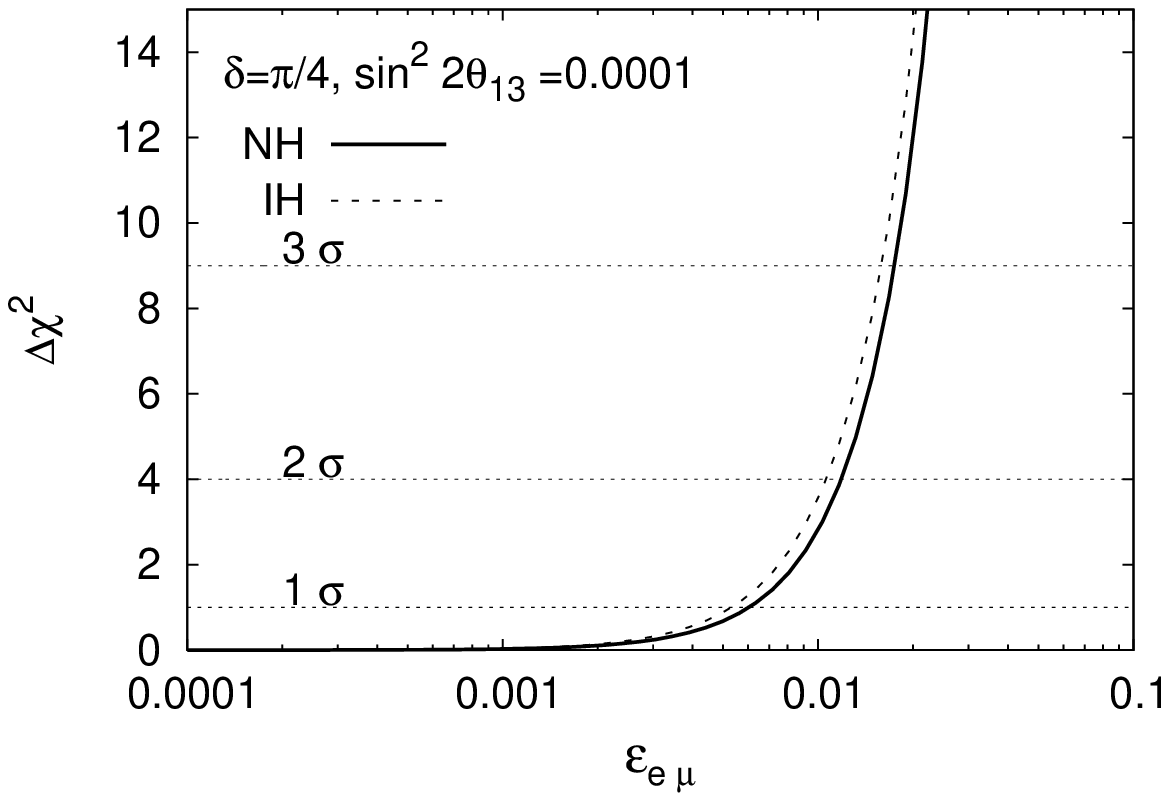,width=0.35\linewidth,clip=}
\end{tabular}
\caption[] {{\small Discovery limits of NSI ($\e_{e\mu}$) for different fixed values of $\theta_{13}$ and $\delta$ considering muon energy to be 5 GeV.}}
\label{fig:emfixnsi5}
\end{figure}

\begin{figure}[h!]
\centering
\begin{tabular}{cc}
\epsfig{file=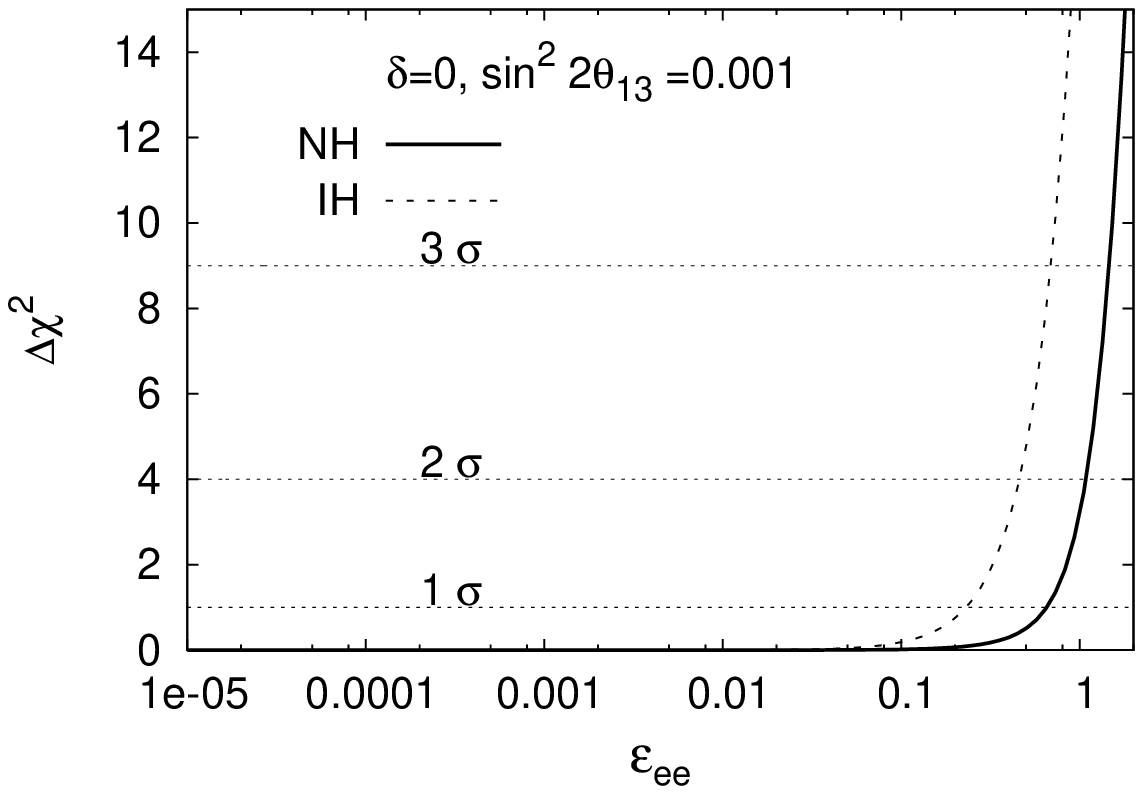,width=0.35\linewidth,clip=}&
\epsfig{file=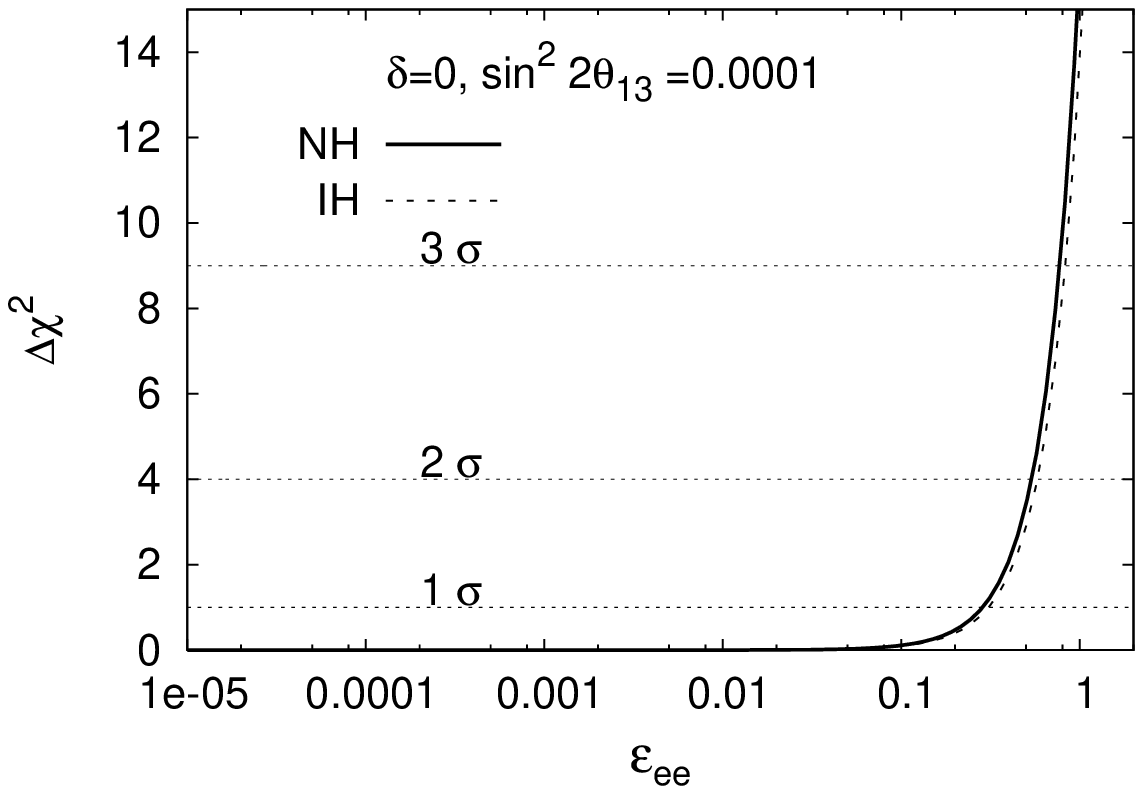,width=0.35\linewidth,clip=}\\
\epsfig{file=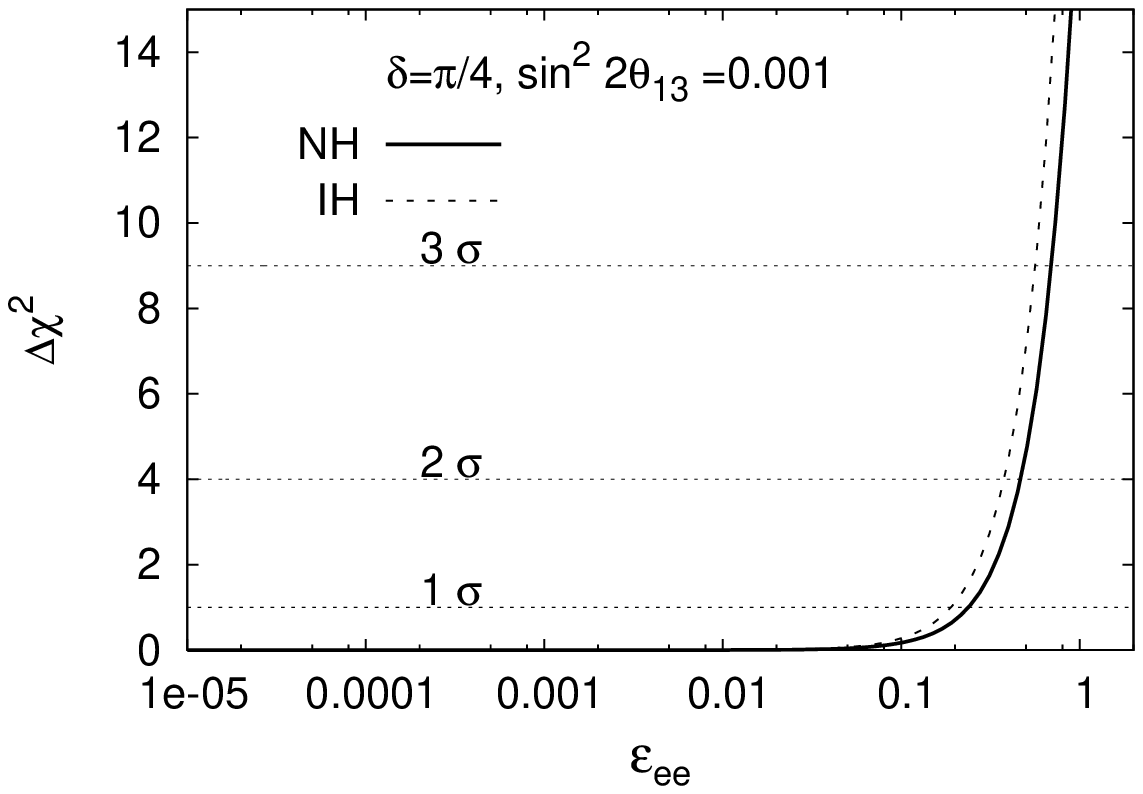,width=0.35\linewidth,clip=}&
\epsfig{file=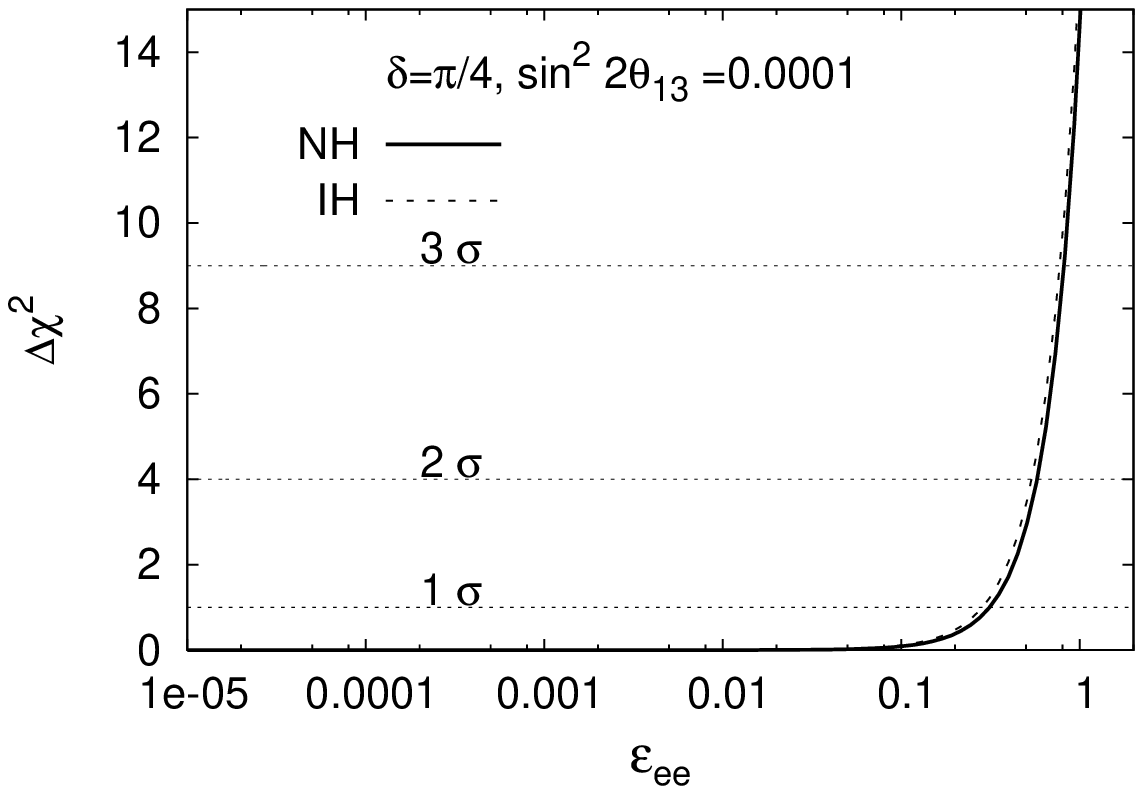,width=0.35\linewidth,clip=}
\end{tabular}
\caption[] {{\small Discovery limits of NSI ($\e_{ee}$) for different fixed values of $\theta_{13}$ and $\delta$ considering muon energy 5 GeV.}}
\label{fig:eensifix}
\end{figure}
\begin{figure}[h!]
\centering
\begin{tabular}{cc}
\epsfig{file=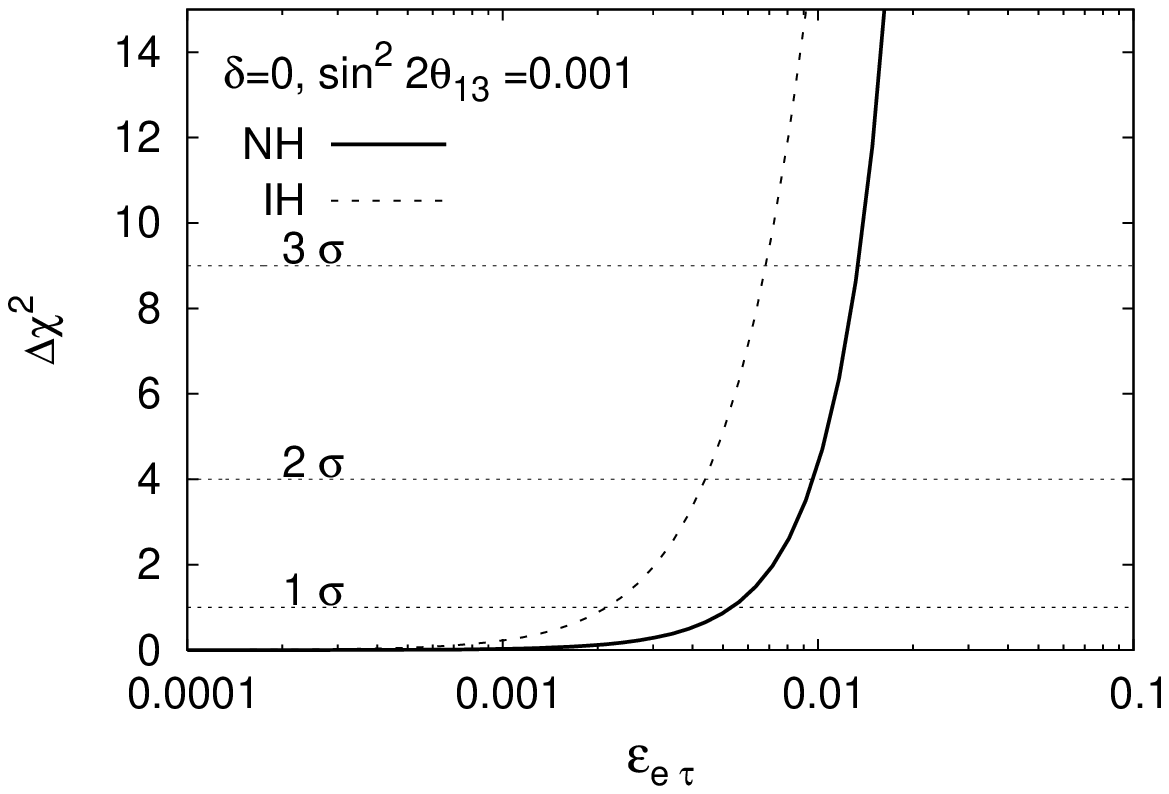,width=0.35\linewidth,clip=}&
\epsfig{file=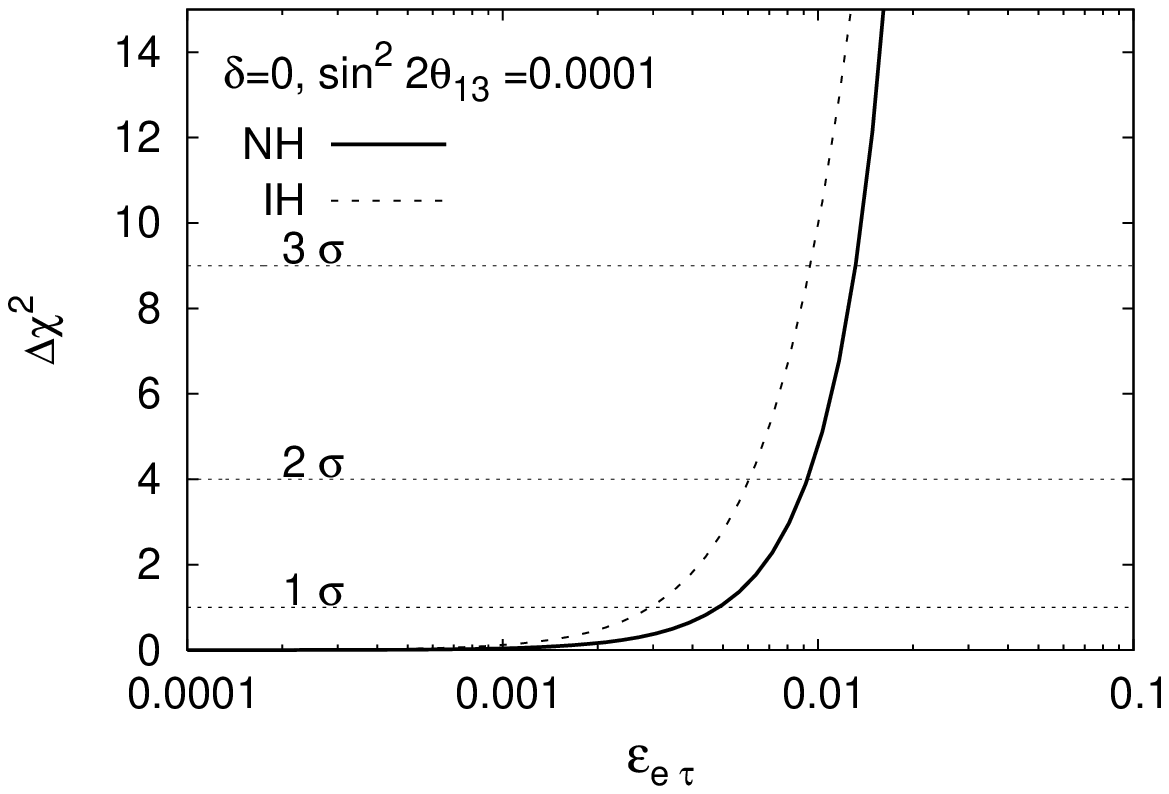,width=0.35\linewidth,clip=}\\
\epsfig{file=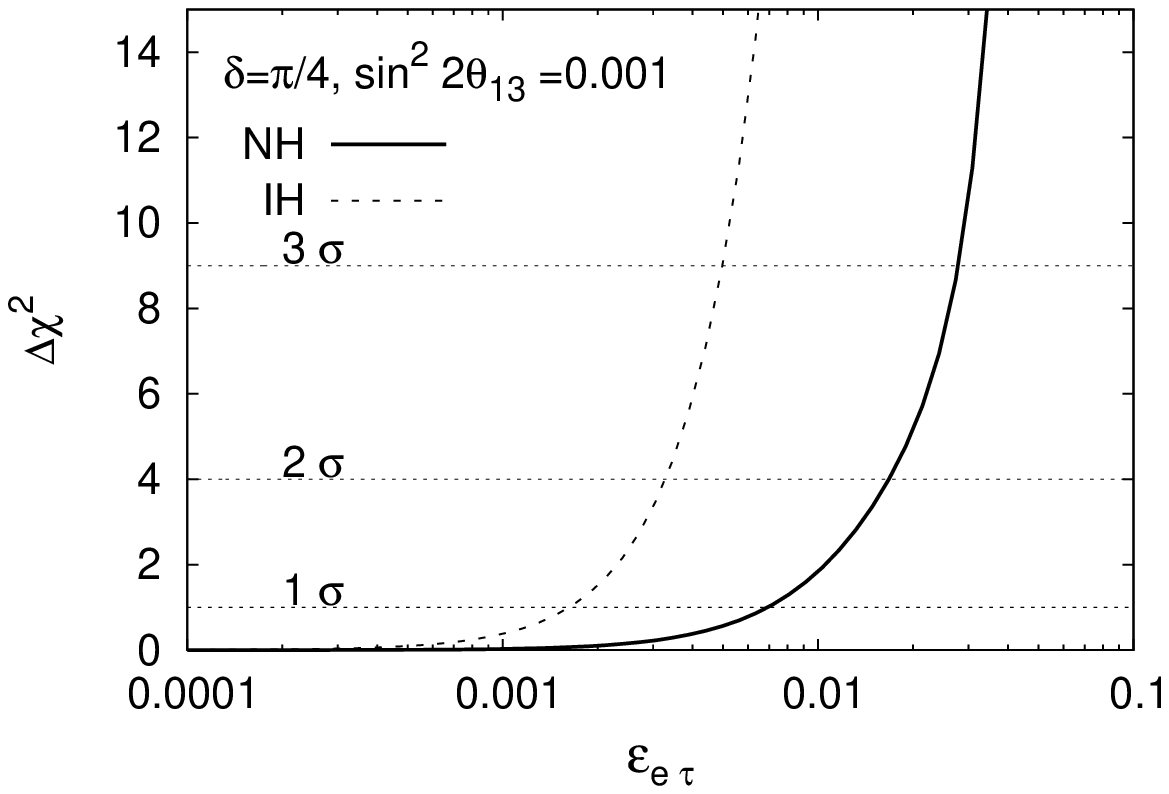,width=0.35\linewidth,clip=}&
\epsfig{file=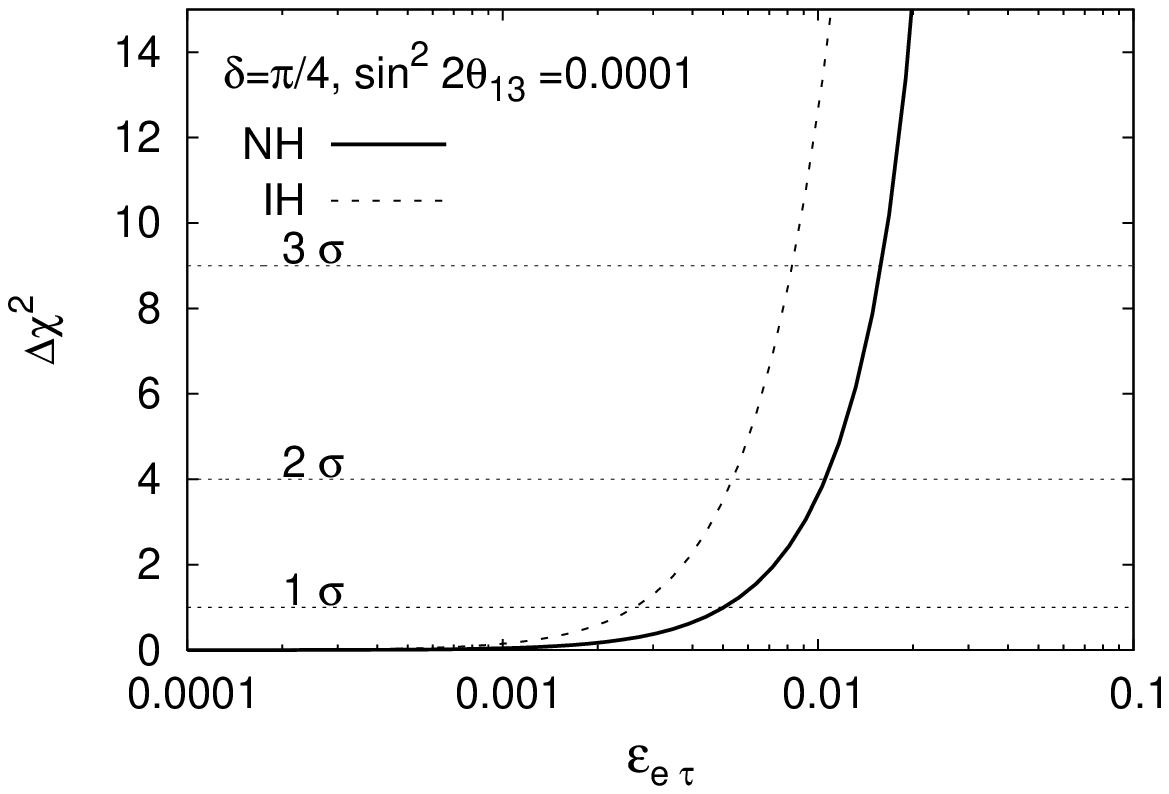,width=0.35\linewidth,clip=}
\end{tabular}
\caption[] {{\small Discovery limits of NSI ($\e_{e\tau}$) for different fixed values of $\theta_{13}$ and $\delta$ considering muon energy 50 GeV.}}
\label{fig:etnsifix50}
\end{figure}

\begin{figure}[h!]
\centering
\begin{tabular}{cc}
\epsfig{file=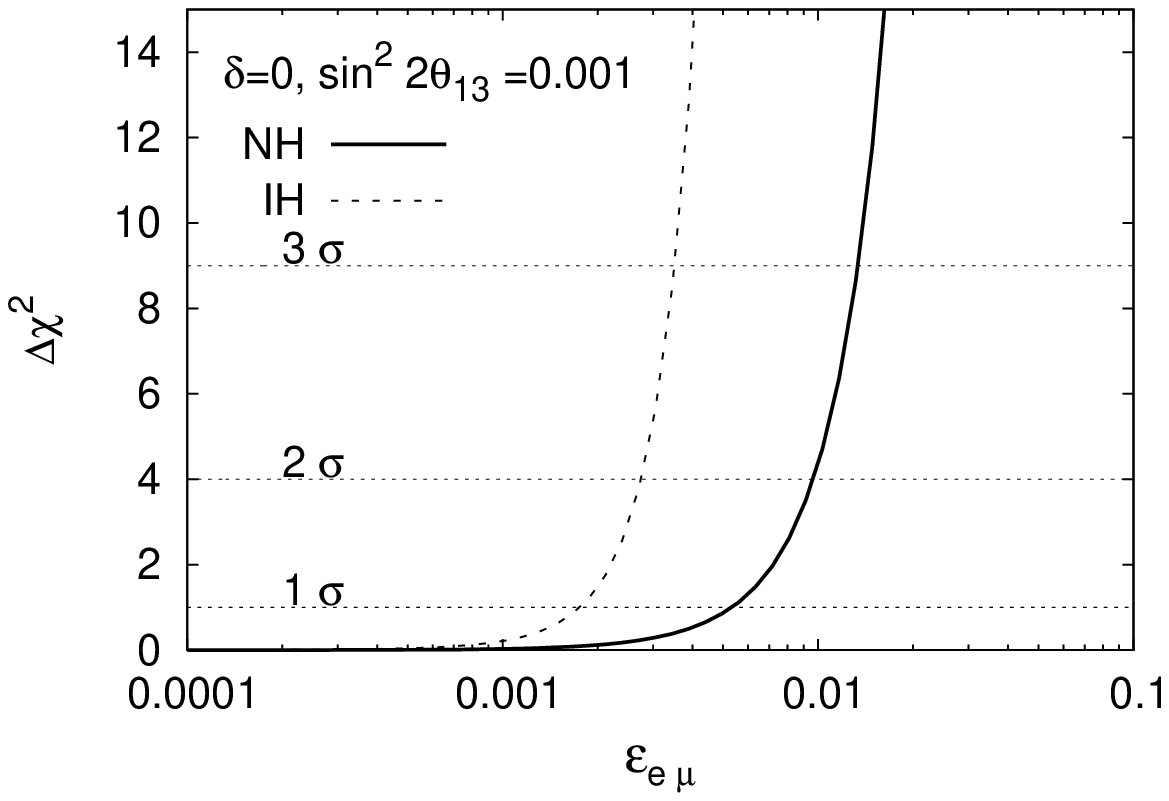,width=0.35\linewidth,clip=}&
\epsfig{file=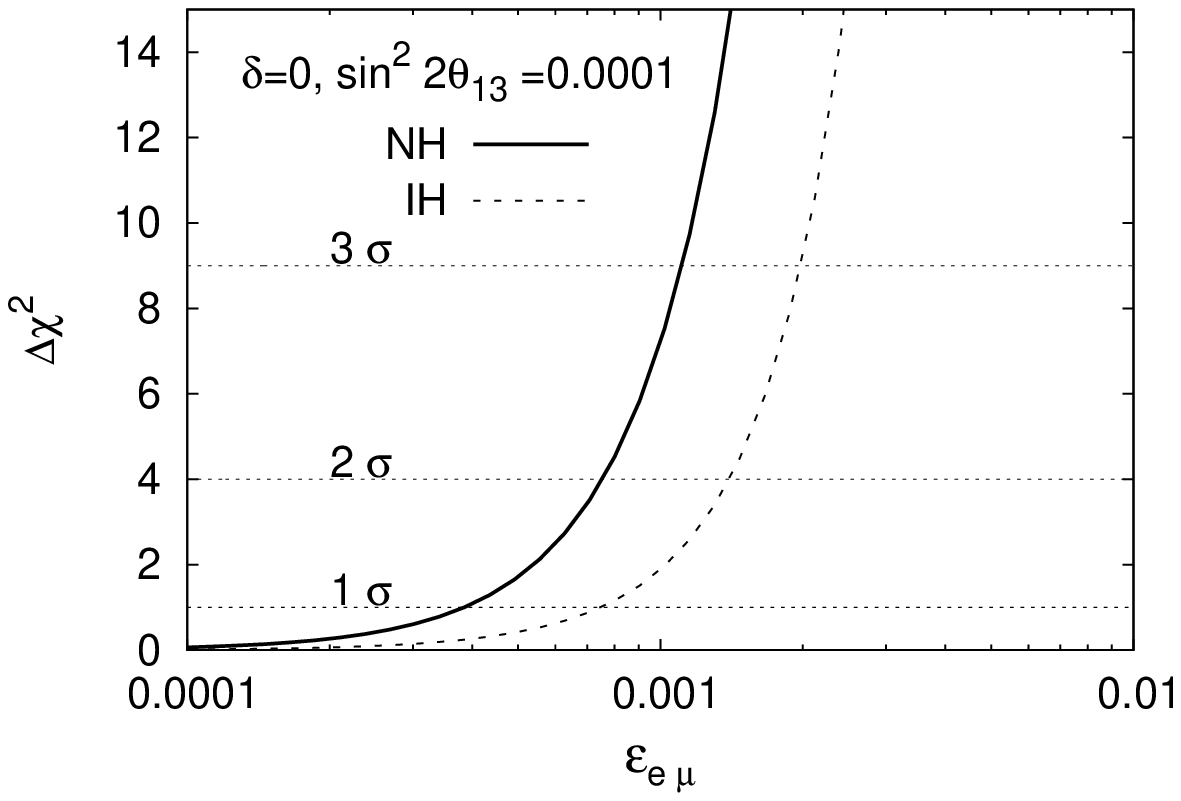,width=0.35\linewidth,clip=}\\
\epsfig{file=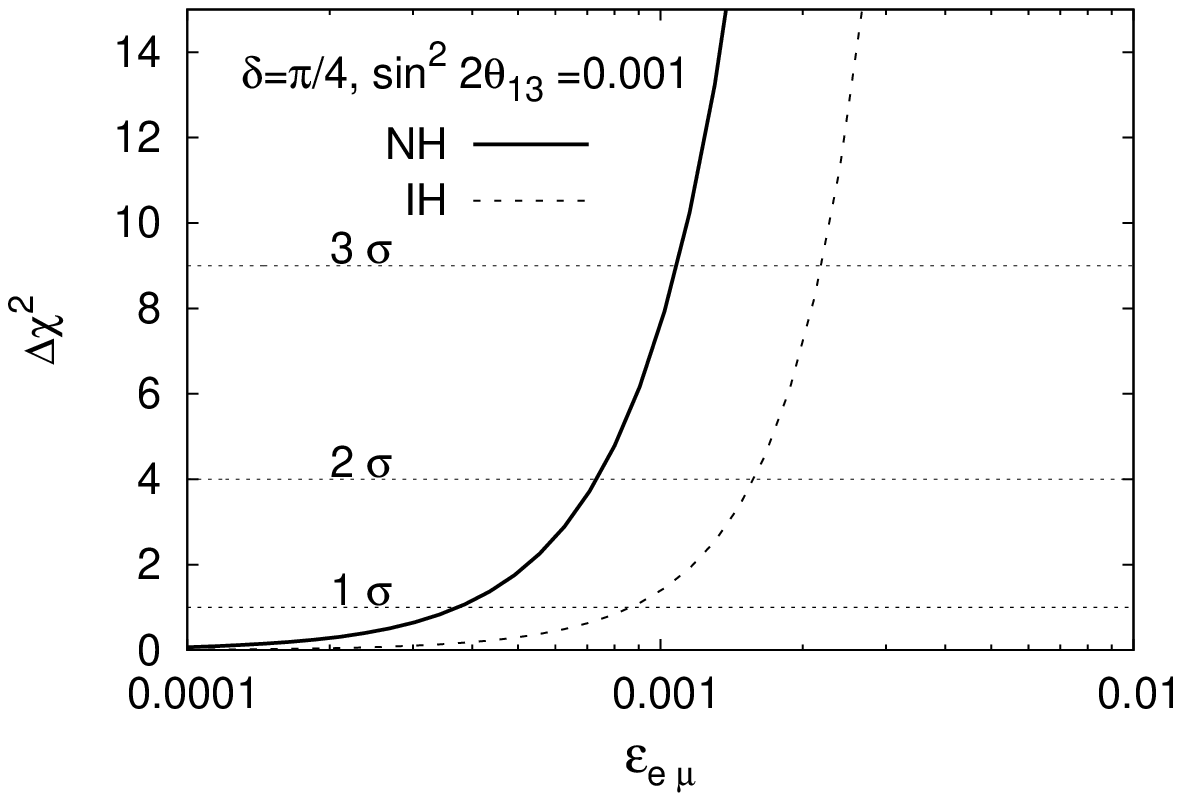,width=0.35\linewidth,clip=}&
\epsfig{file=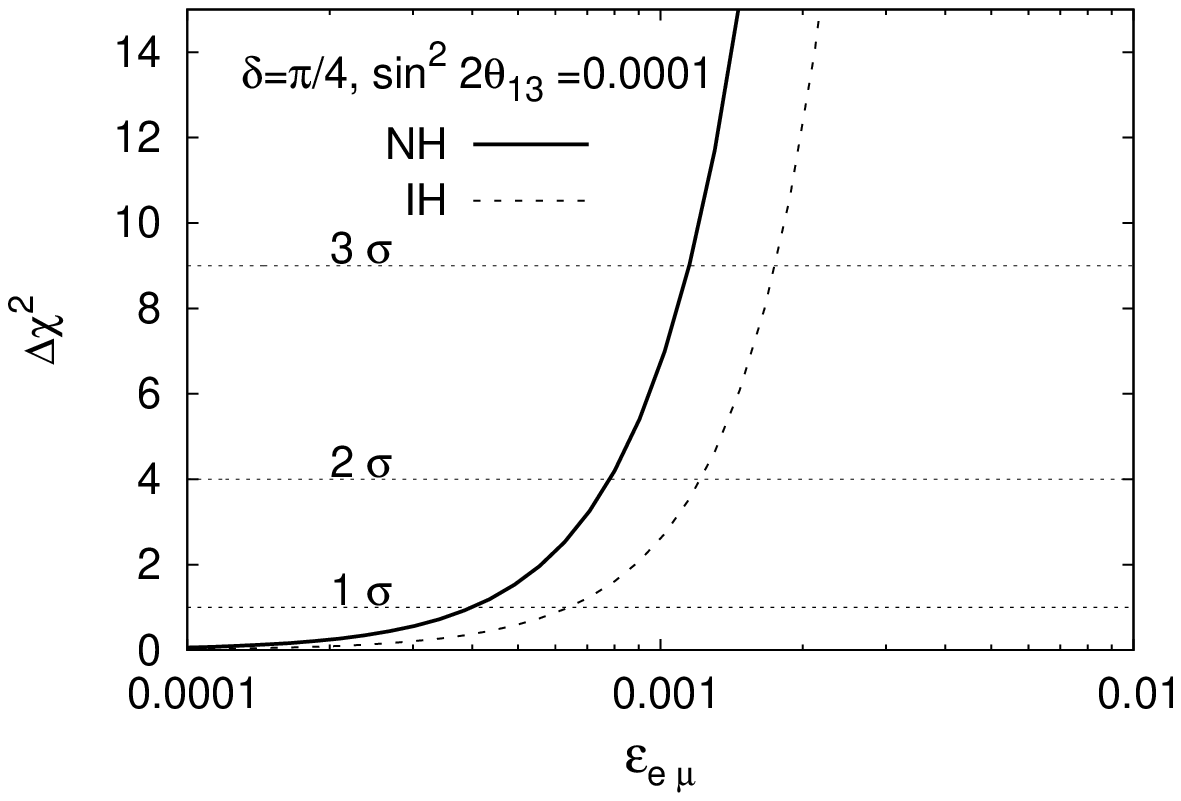,width=0.35\linewidth,clip=}
\end{tabular}
\caption[] {{\small Discovery limits of NSI ($\e_{e\mu}$) for different fixed values of $\theta_{13}$ and $\delta$ considering muon energy 50 Gev.}}
\label{fig:emnsifix50}
\end{figure}

\begin{figure}[h!]
\centering
\begin{tabular}{cc}
\epsfig{file=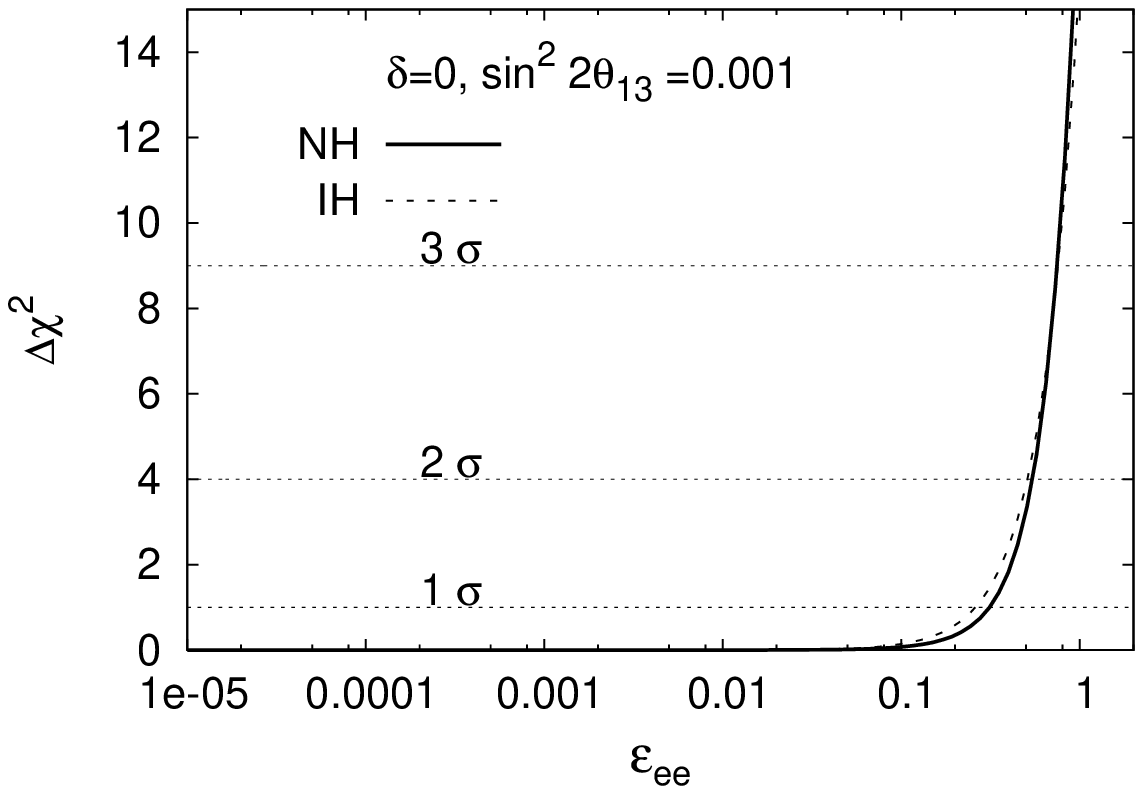,width=0.35\linewidth,clip=}&
\epsfig{file=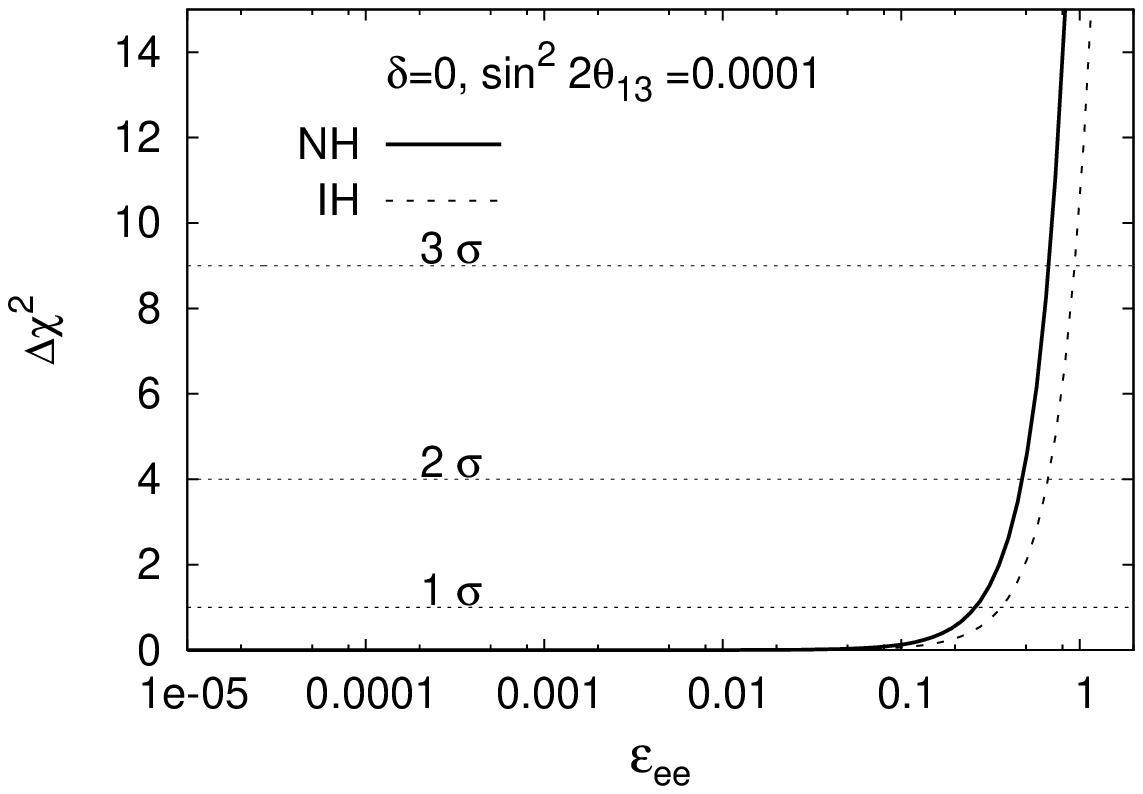,width=0.35\linewidth,clip=}\\
\epsfig{file=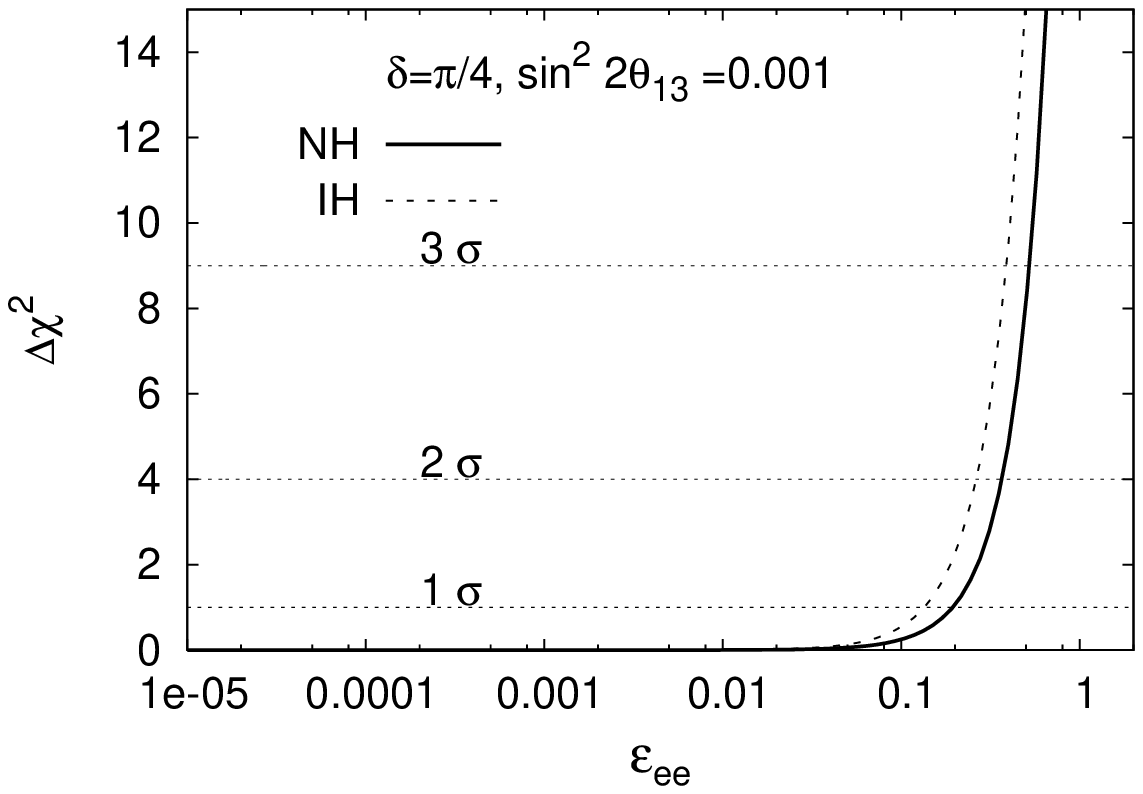,width=0.35\linewidth,clip=}&
\epsfig{file=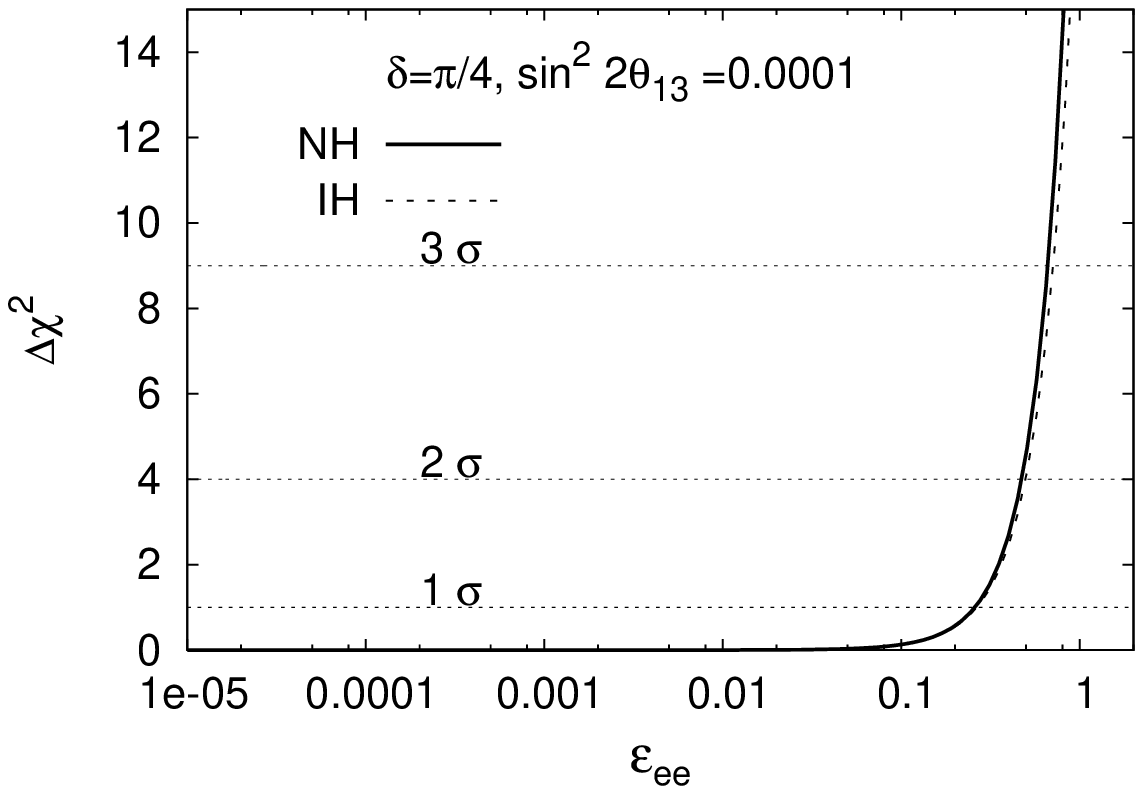,width=0.35\linewidth,clip=}
\end{tabular}
\caption[] {{\small Discovery limits of NSI ($\e_{ee}$) for different fixed values of $\theta_{13}$ and $\delta$ considering muon energy 50 GeV.}}
\label{fig:eensifix50}
\end{figure}

Although we find good discovery limits for hierarchy, $\theta_{13}$ and $CP$ violation in bimagic baseline
even in presence of NSI, however, to get good discovery limits of NSI, the neutrino energy around 5 GeV (as required by magic 
energy conditions) is not appropriate. One can see from the expression of $P(\nu_e \rightarrow \nu_\mu)$ in Eq. (\ref{eq:probcomp}) that the NSI terms are energy independent whereas the terms containing  only vacuum mixing parameters are suppressed by neutrino energy. This feature is present irrespective of specific channel for neutrino oscillation. Naturally for higher
energy the relative effect of NSI parameters are  enhanced with respect to vacuum neutrino mixing parameter and one might
expect to get better discovery liimits of NSIs.  
Now considering 5 GeV as the maximum neutrino energy, we have presented the discovery limits of some NSIs - $\e_{e\tau}$, $\e_{e\mu}$ and $\e_{ee}$ for various fixed values of $\sin^2 \theta_{13}$ and $\delta $ in figures \ref{fig:etnsifix}, \ref{fig:emfixnsi5} and \ref{fig:eensifix} respectively . We can see from these figures that the discovery limits of  $\e_{e\tau}$ and $\e_{e\mu}$  are as low as  ($\approx 0.015$) for either of the heirarchy at $3\sigma$ confidence level. 
For $\e_{ee}$ the limit  is as low as at the order of ($\approx 10^{-1}$) . But at higher neutrino energy say for 50 GeV from figures \ref{fig:etnsifix50} and \ref{fig:emnsifix50} one can see that for IH the discovery limit of $\e_{e\tau}$ can be 
as low as $3 \times 10^{-3}$ and that of $\e_{e\mu}$  could be as low as $7 \times 10^{-4}$. Similarly, for NH the discovery limit of $\e_{e\tau}$ is as low as $0.01$ and for $\e_{e\mu}$  is as low as $0.002$.
 For the case of $\e_{ee}$, from figure \ref{fig:eensifix50} we can see that the discovery limit of the NSI ($\e_{ee}$) is not so good and could be as low as of the order of $10^{-1}$. However, the overall probability of oscillation is suppressed with  the increase in neutrino energy. Naturally it is expected that just increasing energy one may not keep getting better NSI discovery limits. In fact, we have checked at neutrino energy above 60 GeV there is insignificant improvement in discovery limits of 
$\epsilon_{e \mu}$ and $\epsilon_{e\tau}$ in 2540 Km baseline.

\section{Conclusion}
It is found that for getting  good discovery limits for hierarchy, $\sin^2  \theta_{13}$ and $CP$ violation
particularly in the $\nu_e \rightarrow \nu_{\mu}$ oscillation channel, 2540 Km baseline is suitable even when NSI of neutrinos
with matter are present. This is because the  bimagic energies $E_{IH}$ and $E_{NH}$ lie within specific energy range, which is 1-5 GeV for this baseline even in presence
of NSIs with their lower or higher allowed values ( except for $\e_{e\tau} \gsim 0.5$) and this neutrino energy range has been chosen in our analysis with NSIs. It is
important to note that this energy range is also suitable for no NSIs as in that case also bimagic energies are within 1-5 GeV
\cite{dighe}.  

To show what could be the utmost effect to the discovery limits corresponding to no-NSI case, we have considered 
highest possible values as obtained in the model independent case  \cite{nsi1}. However, for model dependent cases \cite{nsi0,adhi,sk} these bounds are  in general, more stringent. 

The discovery limits of hierarchy actually improves in presence of NSIs. Even
one could get discovery limits at $\theta_{13}=0$  for $\e_{e\mu} $ and $\e_{e\tau}$ which in absence of those NSIs are not expected. This is due to the fact that at
bimagic energies the $P_{IH}$ and $P_{NH}$ are unequal even at $\theta_{13}=0$ in presence of those NSIs. This does not
occur for $\e_{ee}$. In this case, as for example, for $\e_{ee} =4$
the  hierarchy discovery limits could be obtained at as low as $\sin^2\theta_{13} \gsim 10^{-4}$. 
Considering highest possible allowed values of $\e_{e\tau}$, $\e_{e\mu}$ and $\e_{ee}$ we find that the
discovery limits of $\sin^2 \theta_{13}$ could be as low as $2 \times 10^{-3}$, $ 6 \times 10^{-4}$ and $2 \times 10^{-5}$ respectively for
normal hierarchy and as low as $2.8 \times 10^{-3}$, $7 \times 10^{-4}$ and $ 1.5 \times 10^{-3}$ respectively for inverted hierarchy. 
Considering favorable values of $\delta $ the discovery limits of $CP$ violation are possible at following $\sin^2 \theta_{13}$ values. For $\e_{e\tau} =3 $ the discovery limits of $CP$ violation could be possible for high value of $\sin^2 \theta_{13}$ at about 0.025 for normal hierarchy only. For inverted hierarchy it is not possible.  
For $\e_{e\mu} =0.33$ the discovery limits of $CP$ violation could be obtained for $\sin^2 \theta_{13}$ as low as $10^{-3}$  
for normal hierarchy and at about $4 \times 10^{-3}$ for inverted hierarchy.
For $\epsilon_{ee} = 4.0$ discovery limits of $CP$ violation cannot be obtained. However, for lower values of both $\e_{e\tau}$ and
$\e_{ee}$ one could get discovery limits of $CP$ violation at some $\sin^2 \theta_{13}$ values.
The discovery limits of NSIs could be improved
if we consider neutrino energy upto 50 GeV and it could be as small as $10^{-3}$ for $\e_{e\mu}$ and $\e_{e\tau}$ and
could be as small as $10^{-1}$ for $\e_{ee}$. These NSI discovery limits essentially would give the upper bound on the respective parameters if they are not discovered. 

It is interesting to note that there are other bimagic baselines with length greater than 6000 Km apart from 2540 Km as discussed before. One may explore the discovery limits of various vacuum neutrino oscillation parameters using those baselines also.

\hspace*{\fill}

\noindent
{\bf  Acknowledgment:} AD likes to thank Council of Scientific and Industrial Research, Govt. of India and ZR likes to thank
University Grants Commission, Govt. of India for providing research fellowships.


\begin{thebibliography}{99}
\bibitem{pmns} K. Hagiwara {\em et al.}, Phys. Rev. {\bf D66},
010001 (2002); B. Pontecorvo Sov. Phys. JETP 26:984 (1968).
\bibitem{pdg} K. Nakamura et al. (Particle Data Group), J. Phys. {\bf G 37}, 075021 (2010) 
and 2011 partial update for the 2012 edition.
\bibitem{ki} K. Kimura, A. Takamura and T. Yoshikawa, hep-ph/0603141;
P. Huber, M. Maltoni and T. Schwetz, Phys. Rev. {\bf D71}, 053006 (2005).
\bibitem{base} G. L. Fogli and E. Lisi, Phys. Rev. {\bf D54}, 3667-3670
(1996); J. Arafune, M. Koike, J. Sato,
Phys. Rev. {\bf D56}, 3093-3099 (1997), Erratum-ibid. {\bf D60}, 119905 (1999);
S. M. Bilenky, C. Giunti and W. Grimus, Phys. Rev. {\bf D58}, 033001 (1998); 
V. D. Barger {\em et al.}, Phys.  Rev.  {\bf D62}, 013004
(2000); M. Freund {\em et al.}, Nucl. Phys.
{\bf B578}, 27-57 (2000); H. Minakata {\em et al.},
Phys.   Rev. {\bf D68}, 033017 (2003), Erratum-ibid. {\bf D70}, 059901
(2004); M. V. Diwan {\em et al.}, Phys. Rev. {\bf D68}, 012002 (2003); D. Choudhury and A. Datta, JHEP {\bf 0507},
058 (2005).
\bibitem{magic1} P. Huber and W. Winter, Phys.  Rev. {\bf D68} , 037301(2003);
P. Huber, J. Phys. {\bf G29}, 1853 (2003); A.Yu. Smirnov,
hep-ph/0610198, A. Asratyan {\em et al.}, hep-ex/0303023, S. K. Agarwalla, S. Choubey and A. Raychaudhuri, Nucl.Phys.
{\bf B771}, 1-27 (2007), S. Choubey {\sl et al}, JHEP {\bf 0912}:020 (2009).
\bibitem{magenergy1} S. K. Raut, R. S. Singh and S.Uma Sankar,Phys.Lett. {\bf B696}, 227-231 (2011).
\bibitem{dighe} A. Dighe, S. Goswami and S. Ray, Phys.Rev.Lett. {\bf 105}, 261802 (2010). 
\bibitem{nsi0} S. Davidson {\em et al.},
JHEP {\bf 0303}, 011 (2003); M. M. Guzzo {\em et al.}, Phys. Lett.
{\bf B591}, 1-6 (2004); J. Barranco {\em et al.}, Phys. Rev.
{\bf D73}, 113001 (2006); G. Mangano {\em et al.}, Phys. {\bf B756},
100-116 (2006); M. Blennow, T. Ohlsson, J. Skrotzki,
Phys. Lett. {\bf B66}, 522-528 (2008); J. Kopp,
M. Lindner, T. Ota, Phys. Rev. {\bf D76}, 013001 (2007);
A. Esteban-Pretel, R. Tomas, J. W. F. Valle, Phys. Rev. {\bf D76}, 053001
(2007); J. Kopp  {\em et al.}, Phys. Rev.
{\bf D77}, 013007 (2008); A. M. Gago {\em et al.}, JHEP {\bf 1001}, 049
(2010); F.J. Escrihuela {\em et al.}, Phys. Rev. {\bf D80},105009 (2009), Erratum-ibid. {\bf D80},
129908 (2009)
; O. Yasuda,
Acta Phys. Polon. {\bf B38}, 3381-3388 (2007).
\bibitem{sk} Super-Kamiokande Collaboration (G. Mitsuka et al.), arXiv:1109.1889 [hep-ex].
\bibitem{adhi} R. Adhikari, S. K. Agarwalla and A. Raychaudhuri, Phys. Lett. {\bf B642}, 111-118 (2006).
\bibitem{val} J. W. F. Valle, arXiv:0608101[hep-ph].
\bibitem{raby} L. J. Hall, V. A. Kostelecky and S. Raby, Nucl. Phys. B267, 415 (1986).
\bibitem{nsi1} C. Biggio, M. Blennow and E. Fernandez-Martinez, JHEP {\bf 0908}, 090 (2009).
\bibitem{perturb} A. Cervera {\em et al.}, Nucl.Phys. {\bf B579}, 17-55 (2000), Erratum-ibid. {\bf B593},
731-732 (2001); M. Freund, Phys. Rev. {\bf D64},
053003 (2001); E. K. Akhmedov {\em et al.}, Nucl.Phys. {\bf B608}, 394-422 (2001);
E. K. Akhmedov {\em et al.}, JHEP {\bf 0404}, 078 (2004).
\bibitem{survprob} T. Kikuchi, H. Minakata  and S. Uchinami,
JHEP {\bf 0903}, 114 (2009).
\bibitem{nsi3} J.Kopp, T. Ota and W.Winter,	Phys. Rev. {\bf D78}, 053007 (2008).
\bibitem{globes1} P. Huber, M. Lindner and W. Winter, Comput. Phys.
Commun.{\bf 167}, 195 (2005); P. Huber, J. Kopp, M. Lind-
ner, M. Rolinec and W. Winter, Comput. Phys. Com-
mun.{\bf 177}, 432 (2007).

\end{thebibliography}
\end{document}